\documentclass[aps,prb,10pt,twocolumn,superscriptaddress,floatfix]{revtex4-1}
\usepackage{chngcntr}
\usepackage{epstopdf}
\usepackage{graphicx}
\usepackage{dcolumn}
\usepackage{bm}
\usepackage{bbm,bbold}
\usepackage{color}
\usepackage{xcolor, soul}
\sethlcolor{green}
\usepackage{amsfonts}
\usepackage{amsmath}
\usepackage{mdframed}
\usepackage[normalem]{ulem}
\usepackage{mathrsfs}   
\usepackage[percent]{overpic}
\usepackage[colorlinks=true,citecolor=blue]{hyperref}
\hypersetup{colorlinks=true,citecolor=blue,linkcolor=blue,urlcolor=blue}
\usepackage{braket}
\usepackage{subfigure}
\begin{document}
\title{Dielectric function and plasmons of doped three-dimensional Luttinger semimetals}
\author{Achille Mauri}
\email{A.Mauri@science.ru.nl}
\affiliation{Scuola Normale Superiore, I-56126 Pisa, Italy}
\thanks{Present address: Radboud University, Institute for Molecules and Materials, NL-6525 AJ Nijmegen, The Netherlands}
\author{Marco Polini}
\affiliation{Istituto Italiano di Tecnologia, Graphene Labs, Via Morego 30, I-16163 Genova, Italy}
\begin{abstract}
Luttinger semimetals are three-dimensional electron systems with a parabolic band touching and an effective total spin $J=3/2$. In this paper, we present an analytical theory of dielectric screening of inversion-symmetric Luttinger semimetals with an arbitrary carrier density and conduction-valence effective mass asymmetry. Assuming a spherical approximation for the single-particle Luttinger Hamiltonian, we determine analytically the dielectric screening function in the random phase approximation for arbitrary values of the wave vector and frequency, the latter in the complex plane. We use this analytical expression to calculate the dispersion relation and Landau damping of the collective modes in the charge sector (i.e.,~plasmons).
\end{abstract}
\maketitle

\section{Introduction}
\label{sect:introduction}
Several three-dimensional (3D) solids are known to present quadratic band touchings at the Fermi energy or in its neighbourhood. Grey tin ($\alpha$-Sn) and mercury telluride (HgTe) were the first materials in which a crossing between parabolic valence and conduction bands was recognized and their rich phenomenology has attracted extensive theoretical and experimental interest for decades~\cite{tsidilkovski, berchenko, averous, gelmont_spu}.

In these solids, the parabolic bands which come into contact are akin to the heavy- and light-hole bands in diamond or zinc-blende semiconductors~\cite{luttinger_pr_1956}, with the difference that one of the two bands has positive concavity due to the presence of strong relativistic effects~\cite{abrikosov_jetp_1971, groves_and_paul_prl_1963}.
A parabolic band touching occurs at the center of the Brillouin zone, is not protected by topology, but, rather, by the cubic symmetry of the crystal.
At $\bm{k} = \bm{0}$, four Bloch wave functions are degenerate and transform under the symmetry group operations as the four spin states of a particle with $J=3/2$. In materials with a centrosymmetric crystal structure, such as $\alpha$-Sn, the conduction and valence electron states in the proximity of the band touching point are described, up to the second order in degenerate $\bm{k}\cdot \bm{p}$ perturbation theory, by the Luttinger Hamiltonian~\cite{luttinger_pr_1956, abrikosov_jetp_1971}. For this reason, 3D semimetals with a parabolic band touching are often referred to as ``Luttinger semimetals" (LSMs). Strictly speaking, in non-centrosymmetric materials such as HgTe, which presents a zinc-blende crystal structure, linear terms in $\bm{k}$ are not ruled out from the $\bm{k}\cdot \bm{p}$ Hamiltonian~\cite{berchenko, gelmont_spu, abrikosov_jetp_1971, qi_revmodphys_2011}. However, linear terms are quantitatively small in HgTe~\cite{berchenko, gelmont_spu, abrikosov_jltp_1975} and deviations from the spectrum of a LSM occur only in a tiny region near the $\Gamma$ point. As a result, treating HgTe as a LSM is adequate for the description of most of its physical properties~\cite{berchenko, gelmont_spu, abrikosov_jltp_1975}.

In LSMs with the Fermi level lying close to the quadratic band touching point electron-electron interactions play an essential role. They have been demonstrated theoretically to drive LSMs either into a strong coupling regime, characterized by scale invariant correlation functions with anomalous exponents for momenta and energies smaller than the effective Bohr scale~\cite{abrikosov_jetp_1971, abrikosov_jltp_1971, abrikosov_jetp_1974, abrikosov_jltp_1975, moon_prl_2013}, or into broken symmetry phases~\cite{janssen_prb_2017, herbut_prl_2014, boettcher_prb_2017, janssen_prb_2016, janssen_prb_2015}. 

Interest in LSMs was revitalized by experimental work on the pyrochlore iridate Pr$_{2}$Ir$_{2}$O$_{7}$, which was found to host a parabolic band touching point at the Fermi level~\cite{kondo_ncomm_2015, cheng_ncomm2017}. In this material, due to the large effective masses of electrons and holes at the quadratic band touching point, correlation effects are important and make it a promising platform for the observation of experimental signatures of the interaction-driven strong-coupling regime~\cite{cheng_ncomm2017}. Effects of interactions in this material, which can lead to anomalous scaling and broken symmetry phases with non-trivial topology, have been the subject of extensive recent investigations~\cite{moon_prl_2013, kondo_ncomm_2015, cheng_ncomm2017, janssen_prb_2017, herbut_prl_2014, boettcher_prb_2017, janssen_prb_2016, janssen_prb_2015, goswami_prb_2017, witczak-krempa_annu_2014}. The LSM model attracted also attention in connection with half-Heusler compounds, such as YPtBi, which were predicted to be topological superconductors~\cite{roy_prb_2019}. In passing, we also mention that LSMs exhibit an extremely rich phenomenology even when electron-electron interactions are neglected, as they can be turned into a plethora of topologically non-trivial states by, e.g.,~applying strain and confinement, or by illuminating them with circularly polarized light~\cite{fu_prb_2007, kondo_ncomm_2015, qi_revmodphys_2011, ghorashi_prb_2018}.

Armed with a low-energy single-particle continuum model Hamiltonian, one can calculate (often analytically) the non-interacting density-density response function $\chi^{(0)}_{nn}({\bm q},\omega)$, which describes the response of the electron system to a spatially varying and time-dependent scalar potential. This quantity was first calculated for a 3D parabolic-band electron system by Lindhard in 1954~\cite{lindhard}. The case of 2D parabolic-band electron systems was analyzed by Stern~\cite{stern_prl_1967}, while the density response function of 2D massless Dirac fermions was first addressed by Shung~\cite{shung} and later analyzed more thoroughly  by the authors of Refs.~\onlinecite{wunsch,hwang,barlas_prl_2007,polini_prb_2008}. More recently, similar calculations have been carried out for three-dimensional Weyl semimetals both in the bulk case~\cite{Lv,zhou,ghosh} and in the presence of Fermi-arc surface states~\cite{andolina_prb_2018}.

This work focuses on the density-density linear response function
$\chi_{nn}(\bm{q},\omega)$ and dynamical screening function $\epsilon(\bm{q},\omega)$ of a LSM. In the celebrated random phase approximation (RPA)~\cite{giuliani_and_vignale}, these are given by Eqs.~(\ref{eq_random_phase_approximation}) and~(\ref{eq_dielectric_function_RPA}) below, respectively. We therefore clearly see that $\chi^{(0)}_{nn}(\bm{q},\omega)$ is a crucial input. 
Screening and dielectric properties of LSMs have been the subject of several studies, which explored the dependence of $\epsilon(\bm{q},\omega)$ on wave vector $\bm{q}$, frequency $\omega$, temperature $T$, and magnetic field~\cite{bailyn_prb_1974,liu_prb_1970, liu_prl_1968, broerman_prb_1970, broerman_prb_1971, bardyszewski_jpcs_1983}. The static wave vector-dependent dielectric function $\epsilon(\bm{q},0)$ was calculated within the RPA in Refs.~\onlinecite{liu_prb_1970, liu_prl_1968, broerman_prb_1970, broerman_prb_1971, bardyszewski_jpcs_1983}. These results were applied to the description of transport properties in presence of scattering by charged impurities~\cite{liu_tosatti, lehoczky_prb_1974, szymanska_pssb_1974}. The RPA dynamic dielectric function, which is relevant for optical properties, was calculated in the long-wavelength $q \rightarrow 0$ limit in Refs.~\onlinecite{sherrington_prl_1968, broerman_prb_1972, boettcher_prb_2019}. In particular, in Ref.~\onlinecite{broerman_prb_1972}, the frequency dependence of the dielectric function was studied at finite $T$ and in presence of lifetime broadening. These results have been recently used to interpret experiments on pyrochlore iridates~\cite{cheng_ncomm2017}. In Ref.~\onlinecite{boettcher_prb_2019}, the dielectric function was calculated for $q \rightarrow 0$ in both the homogeneous and quasi-static limits, including the effects of finite $T$ and band-structure anisotropy, and extending results to the case of a superconducting LSM.

In Ref.~\onlinecite{bardyszewski_jpcs_1983}, the RPA dielectric function was calculated at $T = 0$ for arbitrary $\omega$ and $\bm{q}$, and for arbitrary values of the conduction $m_{\rm c}$ and valence $m_{\rm v}$ effective masses, assuming a spherical approximation. The results in Ref.~\onlinecite{bardyszewski_jpcs_1983}, however, present some inconsistencies. Indeed, the real part of the intra-band dielectric function in Figs. 4 and 5 of Ref.~\onlinecite{bardyszewski_jpcs_1983} exhibits a {\it finite} difference with respect to the Lindhard function of a 3D non-interacting electron gas with a parabolic band in the limit $q \rightarrow 0$. This contrasts with the results presented in this paper, where the two functions are shown to coincide up to leading order in the long-wavelength $q\to 0$ limit. In addition, its imaginary part---Eq. (23) of Ref.~\onlinecite{bardyszewski_jpcs_1983}---deviates from the Lindhard function by terms of order $1/q$ in the limit $q \to 0$, $\omega \to 0$ with $\omega \propto q$, while our results imply a finite difference in this limit.

Several studies were also devoted to the calculation of the screening and dielectric properties of diamond and zincblende semiconductors~\cite{bardyszewski_ssc_1986, yevick_prb_1989, kernreiter_njp_2010, schliemann_epl_2010}, materials in which light- and heavy-hole band states in the neighbourhood of the $\Gamma$ point are described by the Luttinger Hamiltonian, at least when terms linear in $\bm{k}$ in the $\bm{k} \cdot \bm{p}$ electron Hamiltonian can be neglected.

In passing, we note that the dielectric properties of doped LSMs were very recently addressed in Ref.~\onlinecite{tchoumakov_arxiv_2019}, where the wave-vector- and frequency-dependent RPA dielectric function was calculated at $T = 0$, under the assumption of equal conduction and valence band masses, i.e.~$m_{\rm c}=m_{\rm v}$. The resulting dynamically-screened effective electron-electron interaction was then used to calculate the electron self energy.

In this paper, we calculate the density-density response function and the dielectric function of doped LSMs in the normal Fermi liquid state. Following Ref.~\onlinecite{bardyszewski_jpcs_1983}, we work within the so-called spherical approximation of the full Luttinger Hamiltonian (\ref{eq_luttinger_hamiltonian}), see Eq.~(\ref{eq_luttinger_hamiltonian_isotropic}).

A weak-coupling treatment of electron-electron interactions within the RPA~\cite{giuliani_and_vignale} is justified in the Fermi liquid regime when the Fermi energy $E_{\rm F}$ is sufficiently shifted from the quadratic band touching point, i.e.,~when the dimensionless coupling constant $r_{\rm s}$ controlling the relative importance of electron-electron interactions is~\cite{giuliani_and_vignale}
\begin{equation}\label{eq:wigner-seitz}
r_{\rm s} \equiv \left(\frac{9 \pi}{32}\right)^{\frac{1}{3}}\frac{e^2k_{\rm F}/\epsilon_{\rm b}}{\hbar^2 k^2_{\rm F}/(2m)} = \frac{ \left(9 \pi /4\right)^{\frac{1}{3}}}{k_{\rm F} a_{\rm B}} \ll 1~.
\end{equation}
In Eq.~(\ref{eq:wigner-seitz}), $k_{\rm F}$ is the Fermi wave number, $m=m_{\rm c}$ ($m_{\rm v}$) for an $n$-doped ($p$-doped) LSM,  $a_{\rm B}=\epsilon_{\rm b}\hbar^2/(m e^2)$ is the material Bohr radius, and $\epsilon_{\rm b}$ is a background dielectric constant, which arises from the polarization of the bands which are not included in the low-energy model.

We calculate the non-interacting density-density response function $\chi^{(0)}_{nn}(\bm{q},z)$ and the resulting RPA response function $\chi_{nn}(\bm{q},z)$ and dielectric function $\epsilon(\bm{q},z)$ analytically at $T=0$, for arbitrary values of the wave vector $q$ and {\it complex} frequency $z$, allowing $m_{\rm c}\neq m_{\rm v}$. In this sense, the results of this paper for $\chi^{(0)}_{nn}(\bm{q},z)$, $\chi_{nn}(\bm{q},z)$, and $\epsilon(\bm{q},z)$ are more general than those reported in Ref.~\onlinecite{tchoumakov_arxiv_2019}. However, contrary to Ref.~\onlinecite{tchoumakov_arxiv_2019}, we do not present here any calculation of the one-body self-energy and spectral function.

The complete dielectric function constitutes an important tool to describe the properties of the electron system~\cite{giuliani_and_vignale}, because it encodes screening of Coulomb interactions, which is relevant for resistivity calculations and quasiparticle properties, and the dispersion relation of plasmons, which is determined by the zeros of the dielectric function. We derive expressions in a form in which the analytic properties of $\chi^{(0)}_{nn}(\bm{q},\omega)$ in the complex frequency plane are manifest. These expressions are particularly convenient, for example, in diagrammatic calculations, because they can be immediately continued to the Matsubara axis. The knowledge of the dielectric function at arbitrary wave vectors and frequencies allows the determination of the plasmon dispersion relation. We find that a doped LSM supports long-lived plasma excitations, Landau damping setting in, as usual, only at a finite critical wave vector $q_{\rm c}$. The dispersion relation for arbitrary values of $q$ is found numerically, by identifying the maxima of the loss function ${\cal L}(\bm{q},\omega) \equiv - {\rm Im}[1/\epsilon(\bm{q},\omega)]$, which can be measured in electron energy loss spectroscopy~\cite{eels}. Plasmons of an {\it undoped}  LSM ($E_{\rm F}=0$, i.e.~Fermi energy located at the quadratic band touching point) at finite $T$ have been recently calculated in Ref.~\onlinecite{mandal_aop_2019}.

This paper is organized as following. In Sect.~\ref{sect:density_response} we first introduce the continuum-model Hamiltonian we have used to describe LSMs and then introduce the RPA approximation for the density-density response function and the dynamical dielectric screening function. This is the most important section of the paper, as the reader can find in it analytical expressions for the density-density response function of a non-interacting LSM at arbitrary wave vectors $\bm{q}$ and complex frequencies $z$, electron doping $n$, and mass imbalance $\alpha=m_{\rm v}/m_{\rm c}$ between valence and conduction bands.  In Sect.~\ref{sect:plasmons}, we report our results for plasmons and Landau damping. A brief summary of our main results is presented in Sect.~\ref{sect:summary}. Finally, useful technical details on the calculation of the density-density response function of non-interacting LSMs are reported in Appendix~\ref{appendix:appendix1}, while in Appendix~\ref{appendix:appendix2} we discuss the case of hole doping.

\section{Model Hamiltonian, density-density response function, and dynamical dielectric screening function}
\label{sect:density_response}
LSMs display a fourfold degenerate band crossing at the center $\Gamma$ of the Brillouin zone. In the proximity of the crossing point, the single-particle states are described by the effective Luttinger Hamiltonian~\cite{luttinger_pr_1956, abrikosov_jetp_1971}
\begin{eqnarray}
\label{eq_luttinger_hamiltonian}
{\cal H}^{(\rm L)}({\bm k}) &=& \frac{\hbar^{2}}{2 m_{\rm e}}\Bigg[\left(\gamma_{1} + \frac{5}{2} \gamma_{2}\right) k^{2} - 2 \gamma_{3} \left(\bm{k} \cdot \bm{j}\right)^{2} \nonumber \\ &+& 2 \left(\gamma_{3} - \gamma_{2}\right)\left(k_{x}^{2}j_{x}^{2}+k_{y}^{2}j_{y}^{2}+k_{z}^{2}j_{z}^{2}\right)\Bigg]~.
\end{eqnarray}
Here, $\bm{k}$ is the crystal momentum, which, in the spirit of a continuum-model description, will be upgraded to real momentum below, $\bm{j}$ denotes the angular momentum operator in the $J=3/2$ representation, $m_{\rm e}$ is the bare electron mass in vacuum, and $\gamma_{1}$, $\gamma_{2}$, $\gamma_{3}$ are the so-called Luttinger parameters.

The effective Luttinger Hamiltonian describes the heavy- and light-hole valence bands in semiconductors with diamond lattice structure, such as silicon and germanium~\cite{luttinger_pr_1956}. Its form is dictated by symmetry and it is, therefore, very general. The occurrence of a fourfold degenerate set of Bloch wavefunctions at ${\bm k}={\bm 0}$ is related to the existence of four-dimensional irreducible representations of cubic point groups~\cite{abrikosov_jetp_1971}. Eq.~(\ref{eq_luttinger_hamiltonian}) is the most general form of perturbative $\bm{k}\cdot\bm{p}$ Hamiltonian, allowed by cubic symmetry, inversion symmetry and time-reversal invariance~\cite{luttinger_pr_1956, abrikosov_jetp_1971}.

The effective Luttinger Hamiltonian ${\cal H}^{(\rm L)}({\bm k})$ can also be applied to LSMs such as $\alpha$-Sn and Pr$_{2}$Ir$_{2}$O$_{7}$ but with parameters $\gamma_{1}$, $\gamma_{2}$, and $\gamma_{3}$ such that the band structure displays a parabolic node between a valence and a conduction band. In non-centrosymmetric materials such as HgTe and the half-Heusler compound YPtBi, the $\bm{k} \cdot \bm{p}$ Hamiltonian presents additional terms, linear in $\bm{k}$, which are not ruled out due to the absence of inversion symmetry~\cite{boettcher_prb_2019, berchenko, gelmont_spu,  abrikosov_jetp_1971}. In this work, we assume the inversion-symmetric Hamiltonian defined in Eq.~\eqref{eq_luttinger_hamiltonian}. 

In the following, we will carry out calculations only in the spherically-symmetric limit, where the last term in the Luttinger Hamiltonian (\ref{eq_luttinger_hamiltonian}), which violates full rotational invariance (but respects cubic symmetry), is neglected. In this limit, $\gamma_{2} = \gamma_{3} = \gamma$ and the Luttinger Hamiltonian reduces to
\begin{equation}
\label{eq_luttinger_hamiltonian_isotropic}
{\cal H}^{(\rm LS)}({\bm k}) = \frac{\hbar^{2}}{2 m_{\rm e}}\Big[\left(\gamma_{1} + \frac{5}{2} \gamma \right) k^{2} - 2 \gamma \left(\bm{k} \cdot \bm{j}\right)^{2}\Big]~.
\end{equation}
The single-particle eigenstates of ${\cal H}^{(\rm LS)}({\bm k})$ are characterized by definite values of the helicity, i.e. the projection of ${\bm j}$ along ${\bm k}$.

States with helicity $\pm 3/2$ have energy $\left(\gamma_{1} -2\gamma\right) \hbar^{2} k^{2}/(2 m_{\rm e})$, while states with helicity $\pm 1/2$ have energy $\left(\gamma_{1}+ 2\gamma\right) \hbar^{2}k^{2}/(2 m_{\rm e})$. Equation~(\ref{eq_luttinger_hamiltonian}) describes the spectrum of a LSM if $\left|\gamma_{1}\right| < 2\left|\gamma\right|$. We assume, without loss of generality, that $\gamma > 0$. The conduction states then correspond to helicity $\pm 1/2$ and the valence states to helicity $\pm 3/2$.

The second-quantized Hamiltonian of a spherically-symmetric LSM in the presence of Coulomb interactions is
\begin{equation}
\label{eq_many_body_hamiltonian}
\hat{\cal H} = \sum_{\bm{k},\,\alpha,\,\beta} {\cal H}^{(\rm LS)}_{\alpha \beta}\left(\bm{k}\right) \hat{a}^{\dagger}_{\bm{k}, \alpha} \hat{a}_{\bm{k}, \beta}+{\cal H}^{(\rm ee)}~,
\end{equation}
where
\begin{equation}
\label{eq_coulomb_interaction_operator}
\hat{\cal H}^{(\rm ee)} = \frac{1}{2} \sum_{\bm{q}\neq 0} \sum_{\bm{k},\bm{k}^{\prime},\alpha,\beta}v_{\bm{q}} \hat{a}^{\dagger}_{\bm{k}+\bm{q}, \alpha}\hat{a}^{\dagger}_{\bm{k}^{\prime}-\bm{q}, \beta}\hat{a}_{\bm{k}^{\prime}, \beta}\hat{a}_{\bm{k},\alpha}~.
\end{equation}
Here, $\hat{a}^{\dagger}_{\bm{k}, \alpha}$ ($a_{\bm{k}, \alpha}$) creates (annihilates) an electron with momentum $\hbar\bm{k}$ and spin projection $\alpha = -3/2,-1/2,1/2,3/2$ along a fixed quantization axis, independent of $\bm{k}$, and
\begin{equation}\label{eq:Coulomb}
v_{\bm{q}} = \frac{4 \pi e^2}{\epsilon_{\rm b} q^2}~,
\end{equation}
is the 3D Fourier transform of the Coulomb potential~\cite{giuliani_and_vignale}. 

The non-interacting density-density response function~\cite{giuliani_and_vignale} at zero temperature is given, for complex values of the frequency $z$, by:
\begin{equation}
\label{eq_kubo_formula}
\chi_{nn}^{(0)}\left(\bm{q},z \right) =    \frac{1}{V} \sum_{\bm{k},\nu,\nu^{\prime}} \frac{n_{\bm{k},\nu}-n_{\bm{k}+\bm{q},\nu^{\prime}}}{\hbar z+\epsilon_{\bm{k},\nu}-\epsilon_{\bm{k}+\bm{q},\nu^{\prime}}}{\cal F}_{\nu,\nu^{\prime}}(\bm{k},\bm{k}+\bm{q})~.
\end{equation}
Here, $V$ is the 3D electron system volume, $\nu, \nu^\prime = \pm 1/2,\pm 3/2$ are band indices, $n_{\bm{k}, \nu}$ denote the usual Fermi-step occupation numbers at zero temperature, and $\epsilon_{\bm{k},\nu}$ are the single-particle energies, i.e.~$\epsilon_{\bm{k},\pm 1/2} = \hbar^{2} k^{2}/(2 m_{\rm c})$ for the conduction band and $\epsilon_{\bm{k},\pm 3/2} = -\hbar^{2}k^{2}/(2 m_{\rm v})$ for the valence band. Here, $m_{\rm v} = m_{\rm e}/(2 \gamma - \gamma_{1})>0$ and $m_{\rm c} = m_{\rm e}/(2 \gamma + \gamma_{1})>0$, where $\gamma_{1}$ and $\gamma$ have been introduced in Eq.~(\ref{eq_luttinger_hamiltonian_isotropic}). Finally,
\begin{equation}
\label{eq_form_factors}
\begin{split}
{\cal F}_{\nu, \nu^{\prime}}(\bm{k},\bm{k}+\bm{q}) \equiv & \left|\braket{\bm{k},\nu|e^{-i\bm{q}\cdot \bm{r}}|\bm{k}+\bm{q},\nu^{\prime}}\right|^{2} \\ &= \left|D^{(\frac{3}{2})}_{\nu \nu^{\prime}}(\theta_{\bm{k}, \bm{k}+\bm{q}})\right|^{2} ~,
\end{split}
\end{equation}
where $\ket{\bm{k}, \nu}$ denote single-particle eigenstates with momentum $\hbar\bm{k}$ and spin projection $\nu$ along the direction of propagation $\bm{k}$, and $D^{(\frac{3}{2})}_{\nu \nu^{\prime}}(\theta_{\bm{k}, \bm{k}+\bm{q}})$ denotes the unitary rotation matrix of angle $\theta_{\bm{k}, \bm{k}+\bm{q}}$ between the vectors $\bm{k}$ and $\bm{k}+\bm{q}$, in the $J=3/2$ representation. Explicitly, the matrix elements are:
\begin{equation}\label{eq_rotation_matrix_3_2}
 \left|D^{(\frac{3}{2})}_{\nu \nu^{\prime}}(\theta)\right|^{2} =
\begin{bmatrix}
 c^{6} & 3 s^{2} c^{4} & 3 c^{2} s^{4}  & s^{6} \\
  3 s^{2} c^{4}  & (1 - 3s^{2})^{2} c^{2} & (1 - 3c^{2})^{2} s^{2} & 3 c^{2} s^{4} \\
 3 c^{2} s^{4}  & (1 - 3c^{2})^{2} s^{2} & (1 - 3s^{2})^{2} c^{2} & 3 s^{2} c^{4} \\
 s^{6}  & 3 c^{2} s^{4} & 3 s^{2} c^{4}  & c^{6}
\end{bmatrix}~,
\end{equation}
where $c = \cos\left(\theta/2\right)$ and $s = \sin\left(\theta/2\right)$. Summing over pairs of states with equal and opposite helicity, we obtain
\begin{eqnarray}\label{eq_form_factors_sum_intra}
 \mathcal{A}_{\rm intra}(\theta) &\equiv& \sum_{\nu = \pm \frac{1}{2}}  \sum_{\nu^\prime = \pm \frac{1}{2}}  \left|D^{(\frac{3}{2})}_{\nu \nu^{\prime}}(\theta)\right|^{2}  \nonumber\\
 &=& \sum_{\nu = \pm \frac{3}{2}} \sum_{\nu^\prime  = \pm \frac{3}{2}} \left|D^{(\frac{3}{2})}_{\nu \nu^{\prime}}(\theta)\right|^{2}  \nonumber \\ &=& \frac{3}{2} \cos^{2}(\theta) +\frac{1}{2}~,
\end{eqnarray}
for the intra-band form factor, and 
\begin{equation}\label{eq_form_factors_sum_inter}
\mathcal{A}_{\rm inter}(\theta) \equiv \sum_{\nu = \pm \frac{1}{2}} \sum_{\nu' = \pm \frac{3}{2}}     \left|D^{(\frac{3}{2})}_{\nu \nu^{\prime}}(\theta)\right|^{2}  = \frac{3}{2} \sin^{2}(\theta)
\end{equation}
for the inter-band one. It is useful to notice that, due to the equality of the terms in the first and second line of Eq.~(\ref{eq_form_factors_sum_intra}), the form factor $ \mathcal{A}_{\rm intra}(\theta)$ does not depend on the helicity doublet ($\pm 3/2$ or $\pm 1/2$) of the band. As a consequence, results obtained for positive values of the parameter $\gamma$ in Eq.~(\ref{eq_luttinger_hamiltonian_isotropic}) can be generalized in a straightforward manner to the case $\gamma < 0$, in which conduction states have helicity $\pm 3/2$ and valence states have helicity $\pm 1/2$. In the following, we will  consider the case of an $n$-doped LSM, in which the Fermi energy $E_{\rm F}>0$ lies in conduction band.  As discussed in Appendix~\ref{appendix:appendix2}, the results apply with minor changes to the case of $p$-doped samples with $E_{\rm F}<0$.

It is useful to split Eq.~(\ref{eq_kubo_formula}) into the sum of intra- and inter-band contributions:
\begin{equation}
\chi_{nn}^{(0)}\left(\bm{q},z\right) = \chi_{\rm intra}^{(0)}\left(\bm{q},z\right) + \chi_{\rm inter}^{(0)}\left(\bm{q},z\right)~,
\end{equation}
where
\begin{equation}
\label{eq_kubo_formula_intraband_contributions}
  \chi_{\rm intra}^{(0)}(\bm{q},z) \equiv \frac{1}{V} \sum_{\bm{k}} \mathcal{A}_{\rm intra}\left(\theta_{\bm{k}, \bm{k}+ \bm{q}}\right) \frac{n_{\bm{k}, {\rm c}}-n_{\bm{k}+\bm{q}, {\rm c}}}{\hbar z + \epsilon_{\bm{k}, {\rm c}}-\epsilon_{\bm{k}+\bm{q}, {\rm c}}}
\end{equation}
and
\begin{eqnarray}
\label{eq_kubo_formula_interband_contributions}
\chi_{\rm inter}^{(0)}(\bm{q},z) &\equiv & \frac{1}{V} \sum_{\bm{k}} \mathcal{A}_{\rm inter}\left(\theta_{\bm{k}-\bm{q}, \bm{k}}\right) \frac{1-n_{\bm{k}, {\rm c}}}{\hbar z +\epsilon_{\bm{k}-\bm{q}, {\rm v}} -\epsilon_{\bm{k}, {\rm c}}} \nonumber\\
& +& \left[z \rightarrow -z\right]~.
\end{eqnarray}
Here, $n_{{\bm k}, {\rm c}} = \theta(k_{\rm F}-k)$, where $\theta(x)$ is the usual Heaviside step function, $n_{{\bm k}, {\rm v}}=1$, $\epsilon_{{\bm k}, {\rm c}} = \hbar^2k^2/(2 m_{\rm c})$, and $\epsilon_{{\bm k}, {\rm v}} = -\hbar^2k^2/(2 m_{\rm v})$. Throughout this paper, the symbol ``$\left[z \rightarrow -z\right]$" denotes the expression obtained by reversing the sign of the variable $z$ in all terms preceding it.  In the thermodynamic $V\to \infty$ limit, one can replace as usual $V^{-1}\sum_{\bm k} \to \int {\rm d}^3{\bm k}/(2\pi)^3$. Expressions equivalent to Eqs.~(\ref{eq_kubo_formula_intraband_contributions}) and~(\ref{eq_kubo_formula_interband_contributions}) have been used in Refs.~\onlinecite{bardyszewski_jpcs_1983,liu_prl_1968, broerman_prb_1971}.

We immediately notice that, in the long-wavelength $q\to 0$ limit $\chi_{\rm inter}^{(0)}(\bm{q},z)$ vanishes and $\chi_{\rm intra}^{(0)}(\bm{q},z)$ reduces to the well-known density-density response function of a non interacting parabolic-band 3D electron gas~\cite{giuliani_and_vignale}, i.e.~to the so-called Lindhard function $\chi_{\rm L}(\bm{q},z)$. 

In the RPA, the density-density response function $\chi_{nn}(\bm{q},z)$ of the interacting electron system reads as following~\cite{giuliani_and_vignale}
\begin{equation}\label{eq_random_phase_approximation}
\chi_{nn}(\bm{q},z) = \frac{\chi_{nn}^{(0)}(\bm{q},z)}{1-v_{\bm{q}} \chi_{nn}^{(0)}(\bm{q},z)}~.
\end{equation}
In the same approximation, the wave vector- and frequency-dependent dielectric function is~\cite{giuliani_and_vignale}
\begin{equation}\label{eq_dielectric_function_RPA}
\epsilon(\bm{q}, z) = \epsilon_{\rm b} \left[1- v_{\bm{q}}\chi_{nn}^{(0)}(\bm{q},z)\right]~.
\end{equation}

The density-density response functions defined in Eqs.~(\ref{eq_kubo_formula}) and~(\ref{eq_random_phase_approximation}) and the dielectric function in Eq.~(\ref{eq_dielectric_function_RPA}) display branch-cut singularities for real values of the complex frequency $z$, which correspond to electron-hole excitations. The retarded (i.e.~causal) response functions and the dielectric function for real values of the frequency can be obtained~\cite{giuliani_and_vignale} by replacing $z = \omega + i \eta$ and taking the limit $\eta \rightarrow 0^{+}$.

\subsection{Intra-band contribution to the response function}

We first consider the intra-band contribution (\ref{eq_kubo_formula_intraband_contributions}). It is qualitatively similar to the Lindhard function~\cite{giuliani_and_vignale} $\chi_{\rm L}\left(q, z\right)$, which can be obtained by setting ${\cal A}_{\rm intra}(\theta_{{\bm k}, {\bm k}+{\bm q}})=1$ in Eq.~(\ref{eq_kubo_formula_intraband_contributions}).

We start by calculating the imaginary part of the retarded density response function $\chi_{\rm R, intra}^{(0)}(\bm{q}, \omega) \equiv \chi_{\rm intra}^{(0)}(\bm{q}, \omega + i 0^{+})$, for real values of $\omega$. This reads as following
\begin{widetext}
\begin{eqnarray}\label{eq_im_part_intra}
{\rm Im}[\chi^{(0)}_{\rm R, intra}(\bm{q}, \omega)]  &=& -\pi \int \frac{{\rm d}^{3}\bm{k}}{\left(2\pi\right)^{3}} \left[2-\frac{3}{2}\frac{\left(\bm{k}\times \bm{q}\right)^{2}}{k^{2}\left(\bm{k}+\bm{q}\right)^{2}}\right] \left(n_{\bm{k}, {\rm c}}-n_{\bm{k}+\bm{q}, {\rm c}}\right)\delta\left(\hbar \omega +\epsilon_{\bm{k}, {\rm c}}-\epsilon_{\bm{k}+\bm{q}, {\rm c}}\right) \nonumber\\
&=&  - \frac{\pi}{8} N\left(E_{\rm F}\right) \frac{1}{\bar{q}}\Theta\left(1-\nu_{-}^{2}\right) \left[2\left(1-\nu_{-}^{2}\right)-\frac{3}{2}\frac{\bar{q}^2\nu_+^{2}}{\left(\nu_+^{2}-\nu_{-}^{2}\right)}\ln\left(\frac{1-\nu_{-}^{2}+\nu_{+}^{2}}{\nu_{+}^{2}}\right)- \frac{3}{2}\frac{\bar{q}^2\nu_{-}^{2}}{\left(\nu_{+}^{2}-\nu_{-}^{2}\right)}\ln(\nu_{-}^{2})\right] \nonumber\\
& -& [\omega \rightarrow -\omega]~.
\end{eqnarray}
\end{widetext}
Here
\begin{equation}
\nu_{\pm} \equiv \frac{m_{\rm c}}{\hbar k_{\rm F}}\frac{\omega}{q} \pm \frac{q}{2 k_{\rm F}}~,
\end{equation}
$N\left(E_{\rm F}\right) \equiv m_{\rm c}k_{\rm F}/(\pi^{2} \hbar^{2})$ is the density of states at the Fermi energy, and $\bar{q} \equiv q/k_{\rm F}$. Here, $k_{\rm F}=\sqrt{3\pi^2n}$ is the Fermi wave number, written in terms of the electron density $n$.

Note that ${\rm Im}[\chi^{(0)}_{\rm R, intra}(\bm{q}, \omega)]$ is non-zero only in the regions of the $(q,\omega)$ plane where $-1< \nu_{-}<1$ and $-1< \nu_{+}<1$. Physically, these regions represent the continuum of intra-band electron-hole excitations.

The complete intra-band contribution to the density-density response function can be calculated from Eq.~(\ref{eq_im_part_intra}) through the Kramers-Kronig relation~\cite{giuliani_and_vignale}:
\begin{equation}
\label{eq_spectral_representation}
\chi_{\rm intra}^{(0)}\left(\bm{q},z\right) = \frac{1}{\pi} \int_{-\infty}^{\infty}{\rm d}\omega^{\prime} \frac{{\rm Im}[\chi_{\rm R, intra}^{(0)}(\bm{q},\omega^{\prime})]}{\omega^{\prime}-z}~,
\end{equation}
where the integral over $\omega^{\prime}$ runs over the real frequency axis.

The function $\chi^{(0)}_{\rm intra}(\bm{q},z)$ presents branch cuts on the intervals of the real axis in which ${\rm Im}[\chi^{(0)}_{\rm R, intra}(\bm{q}, \omega)]$ is non-zero. For arbitrary complex values of the frequency $z$, away from the branch-cut singularity we get:
\begin{widetext}
\begin{equation}\label{eq_intra}
\begin{split}
\chi_{\rm intra}^{(0)}(\bm{q}, z) = &\chi_{\rm L}(\bm{q},z)+
\frac{3 N(E_{\rm F})}{32}\Bigg\{\frac{1-\bar{q}^{2}}{\bar{q}}\ln\frac{(1+\bar{q})^2}{(1-\bar{q})^2}-4
+\frac{2\nu_+^2}{\nu_{+}+\nu_{-}}f(0, \nu_{+})-\frac{2\nu_{-}^2}{\nu_{+}+\nu_{-}}f\left(0, -\nu_-\right)
\\ & - \frac{2\nu_{-}^2}{\nu_{+}+\nu_{-}}\left[f(\bar{q}, \nu_{+})-g(\bar{q}, \nu_{+})\right]
+\frac{2\nu_{+}^2}{\nu_{+}+\nu_{-}}\left[f(\bar{q}, -\nu_{-})-g(\bar{q}, -\nu_{-})\right]
\Bigg\}~,
\end{split}
\end{equation}
where $\chi_{\rm L}(\bm{q},z)$ is the ordinary Lindhard function~\cite{giuliani_and_vignale}
\begin{equation}\label{eq_lindhard}
\chi_{\rm L}\left(\bm{q},z\right) = - N\left(E_{\rm F}\right) \left\{\frac{1}{2} + \frac{\left(1-\nu_{-}^{2}\right)}{4 \bar{q}}\ln\left[\frac{\nu_{-}-1}{\nu_{-}+1}\right]-\frac{\left(1-\nu_{+}^{2}\right)}{4\bar{q}}\ln\left[\frac{\nu_{+}-1}{\nu_{+}+1}\right]\right\}~.
\end{equation}
The functions $f$ and $g$ are defined by:
\begin{eqnarray}\label{eq_f}
f(\bar{q}, t) \equiv \int_{-1}^{1} \frac{{\rm d} t^{\prime}}{t^{\prime}-t} \ln[(\bar{q} - t^{\prime})^{2}]&=&
\ln[(1-\bar{q})^{2}] \ln\left(\frac{t-1}{t-\bar{q}}\right) 
-\ln[(1+\bar{q})^2]\ln\left(\frac{t+1}{t-\bar{q}}\right)+2{\rm Li}_{2}\left[\frac{1-\bar{q}}{t-\bar{q}}\right] \nonumber\\
&-& 2{\rm Li}_{2}\left[\frac{-(1+\bar{q})}{t-\bar{q}}\right]
\end{eqnarray}
and
\begin{equation}\label{eq_g}
g(\bar{q}, t) \equiv \int_{-1}^{1}\frac{{\rm d}t^{\prime}}{t^{\prime}-t}\ln\left(1+\bar{q}^2-2 \bar{q} t^{\prime}\right)=
\ln \left(1+\bar{q}^{2}-2 \bar{q} t\right)\ln\left(\frac{t-1}{t+1}\right) 
+{\rm Li}_{2}\left[\frac{-2\bar{q}(1+t)}{1+\bar{q}^2-2\bar{q}t}\right] 
-{\rm Li}_{2}\left[\frac{2\bar{q}(1-t)}{1+\bar{q}^2-2\bar{q}t}\right]~,
\end{equation}
\end{widetext}
where ${\rm Li}_{2}[z]$ is the dilogarithm function, i.e.~
\begin{equation}
{\rm Li}_{2}[z] \equiv -\int_{0}^{z}{\rm d}u\frac{\ln\left(1-u\right)}{u}~.
\end{equation}
In Eqs.~(\ref{eq_intra})-(\ref{eq_lindhard}) $\nu_{\pm}$ are promoted to complex quantities:
\begin{equation}
\nu_{\pm} \equiv \frac{m_{\rm c}}{\hbar k_{\rm F}}\frac{z}{q} \pm \frac{q}{2 k_{\rm F}}~.
\end{equation}
The logarithm is defined in a standard way, with a branch cut on the negative real axis and argument ranging from $-\pi$ to $+\pi$. Accordingly, the dilogarithm function presents a branch cut on the interval $[1, +\infty)$ of the real axis. The auxiliary functions $f$ and $g$ are analytic functions of the complex variable $t$, with branch cuts on the interval of the real axis defined by $-1 < t < 1$. As a result, $\chi_{\rm intra}^{(0)}\left(\bm{q}, z \right)$ displays the analytic structure which is expected from the Lehmann (or exact eigenstate) representation~\cite{giuliani_and_vignale}. It is regular in the whole complex plane $z$, except for the intervals $-1<\nu_{+}<1$ and $-1<\nu_{-}<1$ on the real axis, which correspond to the range of energies for which intraband particle-hole pair excitations of momentum $q$ exist. As the cut is crossed, the real part of $\chi_{\rm intra}^{(0)}$ changes continuously but its imaginary part presents a discontinuous change in sign. The retarded response function is obtained by calculating $\lim_{\eta \rightarrow 0^{+}} \chi_{\rm intra}^{(0)}\left(\bm{q},\omega + i \eta \right)$, in such a way that singularities are avoided from above in the complex plane. Taking this limit generates both the real and imaginary parts of the intra-band density response function and it is therefore unnecessary to keep track of the latter separately through Eq.~(\ref{eq_im_part_intra}).

\begin{figure*}[t]
\centering
\begin{overpic}[width=\columnwidth]{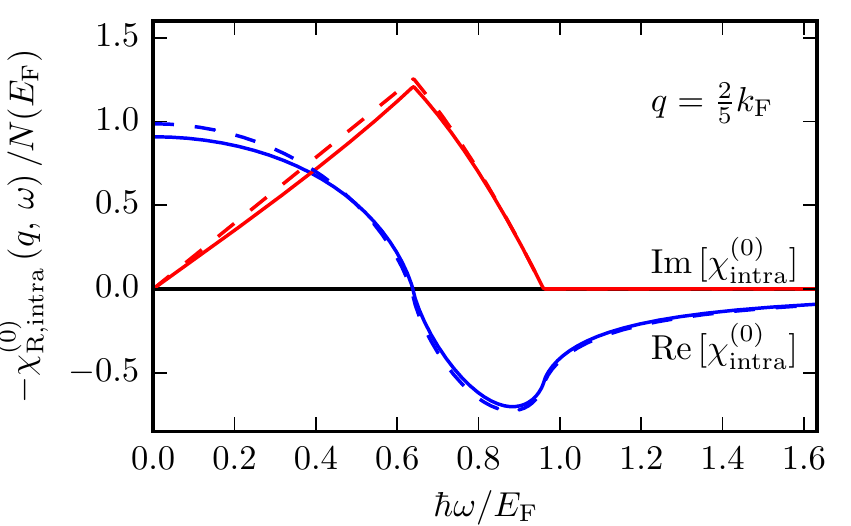}
\put(1,60){(a)}
\end{overpic}
\begin{overpic}[width=\columnwidth]{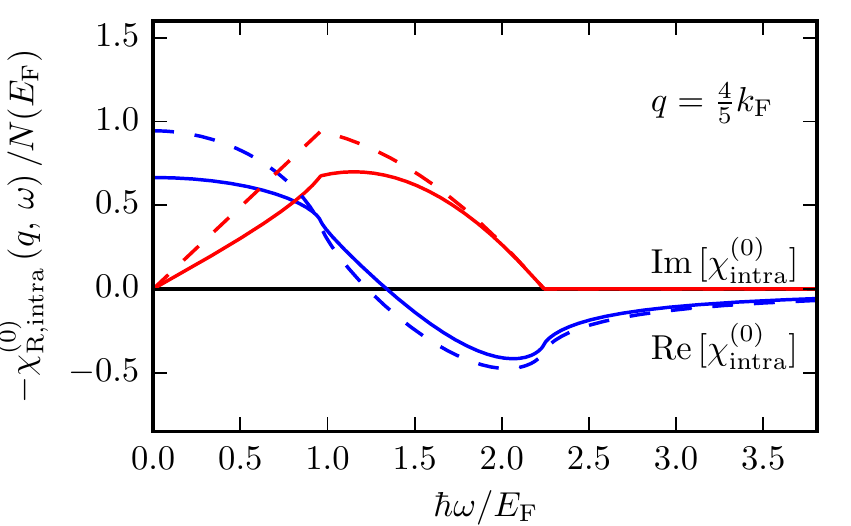}
\put(-2,60){(b)}
\end{overpic}

\begin{overpic}[width=\columnwidth]{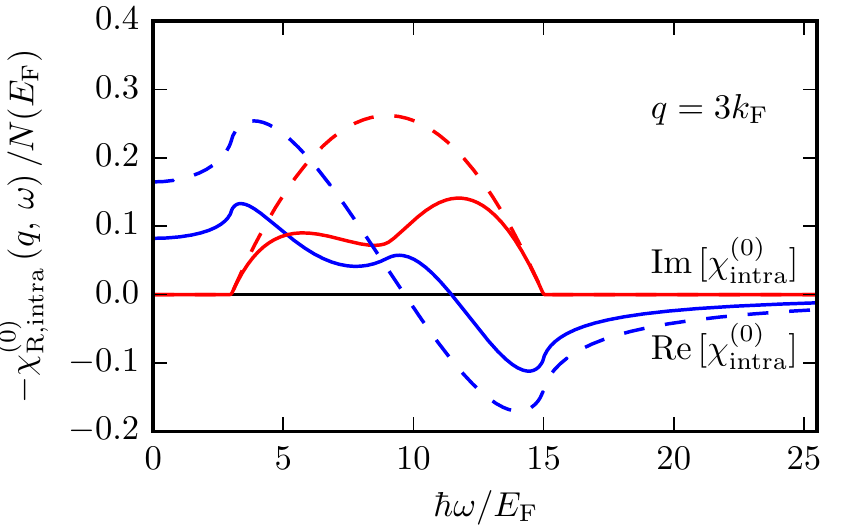}
\put(-2,60){(c)}
\end{overpic}
\begin{overpic}[width=\columnwidth]{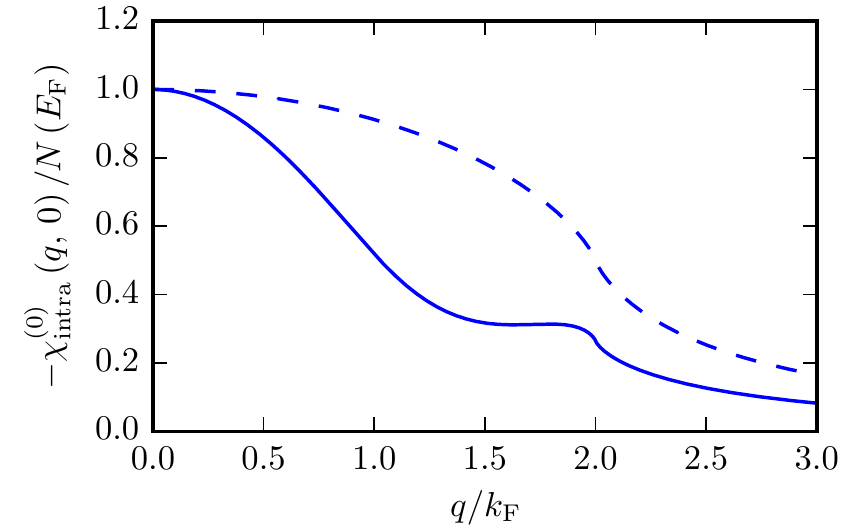}
\put(-2,60){(d)}
\end{overpic}

\caption{%
The intra-band contribution to the non-interacting density-density response function of a Luttinger semimetal (solid lines) is compared with the Lindhard response function~\cite{giuliani_and_vignale} of a 3D system of parabolic-band electrons with mass $m_{\rm c}$ (dashed lines). Blue (red) refers to the real (imaginary) part of $\chi^{(0)}_{\rm R, intra}(\bm{q},\omega)$. All quantities are normalized to the negative of the density of states at the Fermi energy, i.e.~$-N(E_{\rm F})$. (a), (b), and (c) show the dependence of the dynamical response function on the real frequency $\omega$ (in units of $E_{\rm F}/\hbar$) for $q=2k_{\rm F}/5$, $4k_{\rm F}/5$, and $3 k_{\rm F}$, respectively.(d) shows the dependence of the static  response function on $q$ (in units of $k_{\rm F}$). Results in this figure have been obtained by calculating $\chi_{\rm intra}^{(0)}(\bm{q},\omega + i0^{+})$. \label{fig:figure1}
}%
\end{figure*}

Illustrative results on the real frequency axis are shown in Fig.~\ref{fig:figure1}, where the intra-band contribution to the retarded density-density response function is compared with the ordinary Lindhard function~\cite{giuliani_and_vignale} of a 3D system of parabolic-band electrons with mass $m_{\rm c}$. Because the dispersion relation is identical in the two cases, the response functions are qualitatively similar. The imaginary part is nonvanishing in precisely the same regions of the $(q,\omega)$ plane. Differences arise only from the spin $3/2$ structure of the low-energy Luttinger states and these grow with increasing $q$. On the contrary, in the $q\to 0$ limit, $\chi_{\rm intra}^{(0)}(\bm{q},z=\omega+i0^+)$ and $\chi_{\rm L}(\bm{q},z=\omega+i0^+)$ coincide up to leading order in an expansion in powers of $q^{2}$, both at zero frequency, where they approach the famous constant $-N(E_{\rm F})$, and at finite frequency, where they vanish as $nq^{2}/(m_{\rm c}\omega^{2})$, $n$ being the electron density. This is at variance with Figs. 4 and 5 in Ref.~\onlinecite{bardyszewski_jpcs_1983}, where it is seen that $\epsilon_{\rm intra}(\bm{q},z=\omega+i0^+)\neq \epsilon_{\rm L}(\bm{q},z=\omega+i0^+)$ in the $q\to 0$ limit. In Figs.~\ref{fig:figure1}(a)-\ref{fig:figure1}(c), we also see that $\mathrm{Im}[\chi_{\rm R, intra}^{(0)}(\bm{q},\omega)]$ is asymptotically equal to ${\rm Im}[\chi_{\rm L}(\bm{q},\omega)]$ even at finite $\bm{q}$ when $\hbar \omega\to\hbar^{2} (q^{2} + 2 k_{\rm F}q)/(2 m_{\rm c})$, i.e.~the maximum excitation energy of a particle-hole pair of momentum $\bm{q}$. This finding has a simple physical interpretation. For a fixed momentum $\bm{q}$, pairs with the largest possible energy are formed by exciting electrons at the Fermi surface with momentum $\bm{k}$ parallel to the momentum transfer $\bm{q}$. When $\bm{k}$ and $\bm{q}$ are parallel, the non-trivial matrix element due to the spin-$3/2$ structure does not play any role, therefore explaining the asymptotic agreement mentioned above. A similar argument holds for the lower edge in Fig.~\ref{fig:figure1}(c), where $\mathrm{Im}[\chi_{\rm R, intra}^{(0)}(\bm{q},\omega)]$ is seen to be tangent to $\mathrm{Im}[\chi_{\rm L}^{(0)}(\bm{q},\omega)]$ at the minimum excitation energy $\hbar\omega \to \hbar^{2} (q^{2} - 2 k_{\rm F}q)/(2 m_{\rm c})$.

In the static $z \to 0$ limit we obtain, in agreement with Ref.~\onlinecite{bardyszewski_jpcs_1983}:
\begin{equation} \label{eq:chiintra_static}
\begin{split}
 \chi_{\rm intra}^{(0)}(\bm{q}, 0) & = -\frac{3}{4} N\left(E_{\rm F}\right) \Bigg\{\frac{7}{6} + \frac{\bar{q}^{4} - 1}{4 \bar{q}} \ln \left|\frac{1 + \bar{q}}{1 - \bar{q}}\right| \\ & + \frac{\bar{q}}{8} \left(4 - \bar{q}^{2}\right)\left(\frac{4}{3 \bar{q}^{2}} + 1\right) \ln \left|\frac{2 + \bar{q}}{2 - \bar{q}}\right| \\  + \bar{q} \Bigg[ &\Phi\left(-\bar{q}\right) - \Phi(\bar{q})   + \Phi\left(\frac{\bar{q}}{2}\right) - \Phi\left(-\frac{\bar{q}}{2}\right) \Bigg]\Bigg\}~,\\
\end{split}
\end{equation}
where $\Phi\left(x\right)$, for real $x$, is defined as:
\begin{equation}
\begin{split}
\Phi\left(x\right) &  \equiv \lim_{z \to x} {\rm Re}\left[{\rm Li}_{2}\left(z\right)\right] =  - \int_{0}^{x} {\rm d}u \;\frac{\ln\left|1 - u\right|}{u} \\ & = 
\begin{cases}
\mathrm{Li}_{2}\left(x\right) & \text{for } x < 1\\
\frac{\pi^{2}}{3} - \frac{1}{2} \ln^{2}\left(x\right) - \mathrm{Li}_{2}\left(\frac{1}{x}\right) & \text{for } x > 1
\end{cases}~.
\end{split}
\end{equation}
The static function is illustrated in Fig.~\ref{fig:figure1}(d). Comparing the Lindhard function with $\chi_{\rm intra}^{(0)}(\bm{q}, 0)$, we clearly see that the matrix elements (\ref{eq_form_factors}) associated with the non-trivial structure of spin-$3/2$ states lead to a dramatic reduction of the intra-band response of a LSM and strong qualitative modifications in the region of finite $q$. We even see a non-monotonic dependence of $\chi_{\rm intra}^{(0)}(\bm{q},0)$ on $q$ in a range of values of $q$. This strong reduction in polarizability is a consequence of the spin-$3/2$ structure of the Luttinger Hamiltonian.

In the static case, both the Lindhard function and the intra-band response function of a LSM present a singular derivative at $q= 2k_{\rm F}$. This singularity gives rise to Friedel oscillations~\cite{giuliani_and_vignale, gonzales_prb_1989} of period $\pi/k_{\rm F}$ in the density profile induced by a static point-like charge introduced in the 3D electron system. In addition, $\chi_{\rm intra}^{(0)}(\bm{q},0)$ presents a weaker logarithmic singularity in the third derivative at $q = k_{\rm F}$. However, this latter singularity disappears from the total static density response function $\chi_{nn}^{(0)}\left(\bm{q}, 0\right) = \chi_{\rm intra}^{(0)}\left(\bm{q}, 0\right) + \chi_{\rm inter}^{(0)}\left(\bm{q}, 0\right)$, as it can be checked from Eq.~\eqref{eq:chiintra_static} and ~\eqref{eq:chiinter_static} (see also Ref.~\onlinecite{gonzales_prb_1989}).

\begin{figure}[h]
\centering
\begin{overpic}{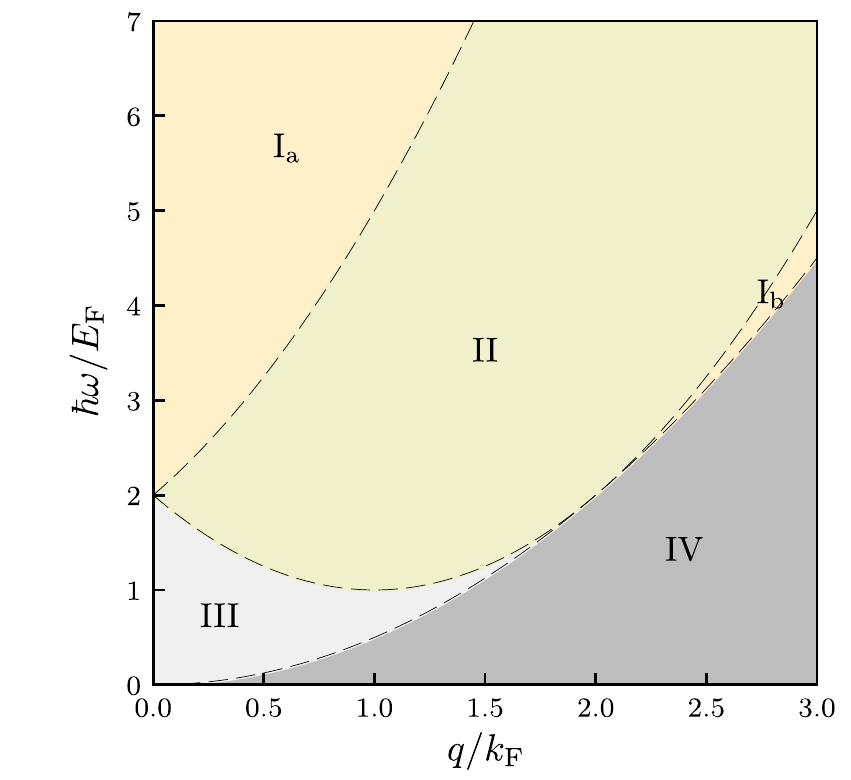}
\end{overpic}
\caption{\label{fig:figure2}
The inter-band particle-hole continuum of a Luttinger semimetal in the $(q,\omega)$ plane. In region ${\rm IV}$, $\omega < \hbar q^{2}/[2(m_{\rm c}+m_{\rm v})]$ ($r^{2}<0$), the existence of inter-band particle-hole pairs is kinematically forbidden. In region ${\rm III}$, $r^{2}>0$ but inter-band transitions are Pauli blocked. This happens for $k_{0} + r < k_{\rm F}$, when the spherical surface of electron momenta lies inside the Fermi sphere. In region ${\rm II}$, $k_{0}+r>k_{\rm F}$ and $\left(k_{\rm 0}-r\right)^{2}<k_{\rm F}^{2}$, implying that the spherical surface overlaps partially with the Fermi sphere. In this region, a fraction of inter-band transitions is blocked by Pauli exclusion principle. In the high-energy or high-momentum regions ${\rm I}_{\rm a}$ and ${\rm I}_{\rm b}$, Pauli blocking has no effect and $k_0-r> k_{\rm F}$ or $k_{0}-r < -k_{\rm F}$. While in the $(k_{0},r)$ parametrization one can distinguish different regions of the inter-band electron-hole continuum without specifying the mass imbalance $\alpha$, in the $(q,\omega)$ plane the boundaries depend on $\alpha$. Results in this figure refer to $\alpha=1$.}
\end{figure}
\begin{figure}[h!]
\centering
\begin{overpic}[width=\columnwidth]{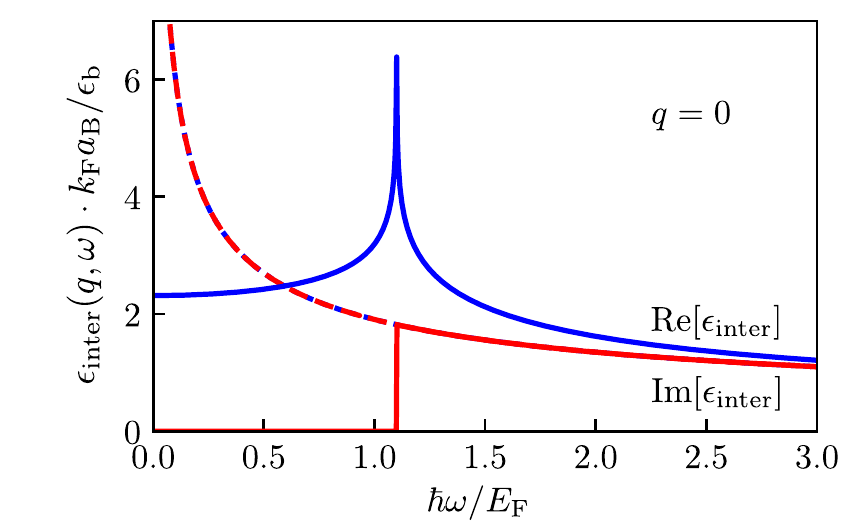}
\put(1,60){(a)}
\end{overpic}
\begin{overpic}[width=\columnwidth]{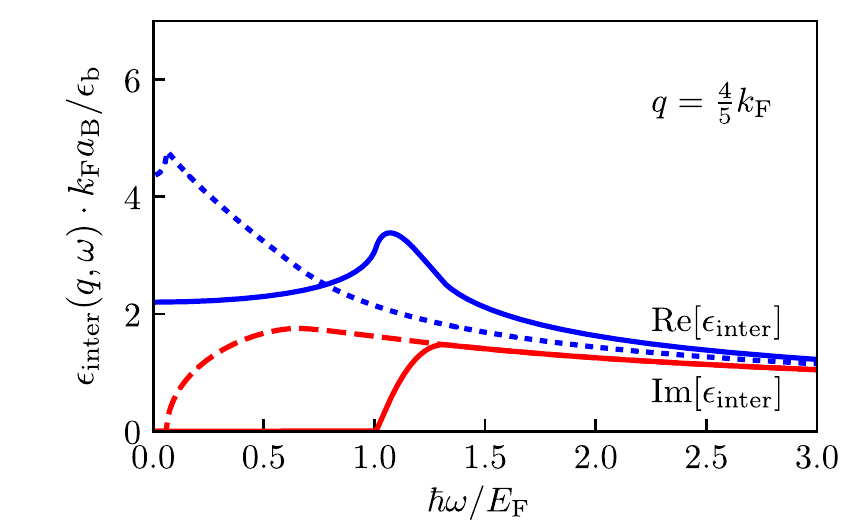}
\put(1,60){(b)}
\end{overpic}
\begin{overpic}[width=\columnwidth]{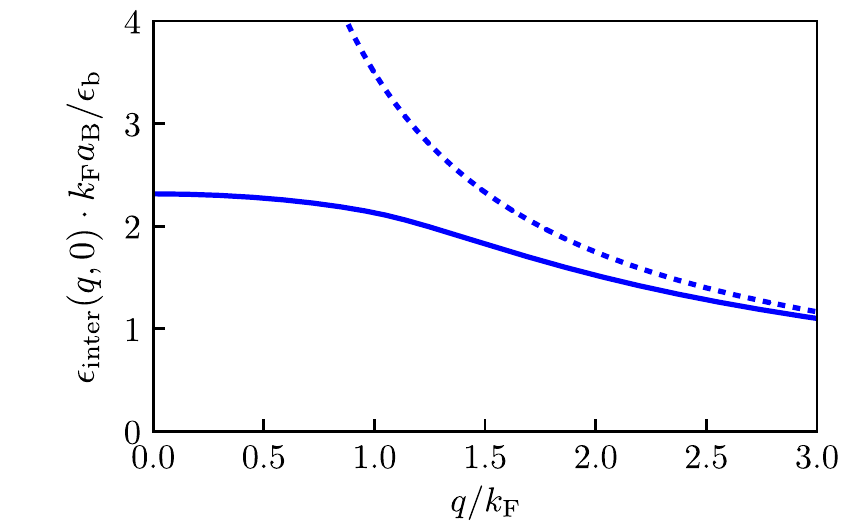}
\put(1,60){(c)}
\end{overpic}
\caption{\label{fig:figure3}
The inter-band contribution (\ref{eq:interband_dielectric}) to the dielectric function (solid lines) of a doped Luttinger semimetal is compared with the dielectric function of an undoped system (dashed and dotted lines). (a) Results for $q=0$. In the undoped case, $\epsilon(\bm{q},\omega)$ diverges as $1/\sqrt{\omega}$ and its real and imaginary parts coincide~\cite{sherrington_prl_1968, broerman_prb_1972, cheng_ncomm2017}. In the doped case, Pauli blocking cuts off this singularity at $\omega^* = E_{\rm F}(1+1/\alpha)/\hbar$, where ${\rm Im}[\epsilon_{\rm inter}(0,\omega)]$ presents a discontinuous jump to zero~\cite{sherrington_prl_1968, broerman_prb_1972, cheng_ncomm2017}. Correspondingly, ${\rm Re}[\epsilon_{\rm inter}(0,\omega^*)]$ displays a logarithmic divergence~\cite{sherrington_prl_1968, broerman_prb_1972, cheng_ncomm2017}. (b) All divergencies disappear~\cite{goettig_pssb_1975} at finite $q$. (c) The static $\omega=0$ case. In the undoped case~\cite{liu_prl_1968}, $\epsilon_{\rm inter}(\bm{q},0)$ diverges as $1/q$. Pauli blocking cures this divergence, leaving a regular, but strongly density-dependent dielectric function~\cite{liu_tosatti, bailyn_prb_1974}. Results in this figure refer to $\alpha = 10$.}
\end{figure}
\subsection{Inter-band contribution to the response function}

The inter-band density-density response $\chi^{(0)}_{\rm inter}(\bm{q},z)$ is particularly interesting because it displays the peculiar behavior associated with the absence of a gap between the valence and conduction band~\cite{liu_prl_1968, sherrington_prl_1968, liu_tosatti, bailyn_prb_1974, broerman_prb_1972, cheng_ncomm2017, averous}.

Technical details on the calculation of $\chi^{(0)}_{\rm inter}(\bm{q},z)$ are reported in Appendix~\ref{appendix:appendix1}. Here, we report only the final results. It is convenient to decompose the inter-band density response function as:
\begin{equation}
\label{eq_interband_response_decomposition}
\chi_{\rm inter}^{(0)}(\bm{q}, z) = \chi_{\rm u}^{(0)}(\bm{q},z) + \delta \chi_{\rm inter}^{(0)}(\bm{q},z)~,
\end{equation}
where $\chi_{\rm u}^{(0)}(\bm{q},z)$ is the density-density response of  an {\it undoped} LSM, in which the Fermi energy lies at the parabolic band touching point, while $\delta \chi_{\rm inter}^{(0)}(\bm{q},z)$ represents the contribution arising from the presence of a finite electron concentration in the conduction band.  

For arbitrary wave vectors and complex frequencies, and arbitrary values of the mass imbalance $\alpha=m_{\rm v}/m_{\rm c}$ we find
\begin{widetext}
\begin{eqnarray}\label{eq_undoped_interband_response_function}
\chi_{\rm u}^{(0)}(\bm{q},z) &=&  -\frac{3 i}{16\pi}\frac{m_{\rm c}+m_{\rm v}}{\hbar^{2}}r+  \frac{3 \left(\alpha + 1\right)}{32 \pi \hbar k_{0} z}\left[\left(r^2-k_{0}^2\right)^2\arctan\left(\frac{ik_{0}}{r}\right)+\alpha^2\left(\frac{r^2}{\alpha^2}-k_{0}^2\right)^2\arctan\left(\frac{i\alpha k_0}{r}\right)\right]\nonumber\\ 
&+& {\left[z\rightarrow-z\right]}
\end{eqnarray}
and
\begin{equation}
\label{eq_doped_interband_response_function}
\begin{split}
\delta \chi_{\rm inter}^{(0)}(\bm{q}, z)&  =  \frac{3 m_{\rm c}k_{\rm F}}{32\pi^2\hbar^{2}}\Bigg\{ -\frac{1}{2\bar{k}_{0}}\left[4\alpha \bar{k}_{0}+\frac{1-\bar{q}^2}{\alpha + 1}\ln \frac{\left(1+\bar{q}\right)^2}{\left(1-\bar{q}\right)^2}\right] +\frac{\alpha+1}{2\bar{k}_{0}}\Bigg[\left((\bar{r}-\bar{k}_{0})^2-1\right)\ln\frac{\bar{r}-\bar{k}_{0}+1}{\bar{r}-\bar{k}_{0}-1}\\ & -\left((\bar{r}+\bar{k}_{0})^2-1\right)\ln\frac{\bar{r}+\bar{k}_{0}+1}{\bar{r}+\bar{k}_{0}-1}\Bigg] - \frac{\left(\bar{z} - \bar{q}^{2}\right)^{2}}{2 \bar{q} \bar{z}}\Bigg[f(-\bar{q}, \bar{r}-\bar{k}_{0})\\ & +f(-\bar{q}, -\bar{r}-\bar{k}_{0})-g\left(-\bar{q},\frac{\bar{r}^2-1-\bar{k}_{0}^2}{2\bar{k}_{0}}\right)\Bigg] + \frac{\left(\alpha \bar{z} -\bar{q}^{2}\right)^{2}}{2 \bar{q} \bar{z}}\Big[f(0,\bar{r}-\bar{k}_{0})\\ & +f(0, -\bar{r}-\bar{k}_{0})\Big] \Bigg\}\\
&+\left[z \rightarrow -z\right]~.
\end{split}
\end{equation}
\end{widetext}
Here, $\bar{z} =  \hbar z/E_{\rm F}$ is the frequency in units of the Fermi energy, 
\begin{equation}
\label{eq_definition_k0}
\bm{k}_{0} = \frac{\bm{q}}{1+\alpha}
\end{equation}
and
\begin{equation}\label{eq_definition_r}
r = i\sqrt{\frac{\alpha}{\alpha+1} \left(\frac{q^{2}}{\alpha+1} -\frac{2 m_{\rm c} z}{\hbar}\right)}
\end{equation}
are auxiliary variables, $\bar{r}\equiv r/k_{\rm F}$, and $\bar{k}_{0}\equiv k_{0}/k_{\rm F}$. (For example, the parameter $\alpha$ is on the order of $10$ for grey tin and mercury telluride.) In Eq.~(\ref{eq_definition_r}), the square root is defined in a standard way, with a branch cut on the negative real axis and a branch choice such that ${\rm Re} \left[\sqrt{x}\right] \geq 0$ for complex $x$. The function $\arctan(x)$ in Eq.~(\ref{eq_undoped_interband_response_function}) is defined in the complex plane as:
\begin{equation}
 \arctan(x) \equiv \frac{i}{2} \ln\left(\frac{i + x}{i - x}\right)~,
\end{equation}
with the standard choice of branch cut for the logarithm. The auxiliary variables $r$ and $k_{0}$ can be ascribed a physical meaning. Indeed, it is easy to check that the possible electron momenta of inter-band particle-hole pairs span a spherical region of the momentum space centered in $\bm{k}_{0}$ and of radius $r$.

Equations~(\ref{eq_undoped_interband_response_function}) and~(\ref{eq_doped_interband_response_function}) define analytic functions in the complex frequency plane. They exhibit branch cuts only on the intervals of the real frequency axis which correspond to the energy of inter-band particle-hole pairs (or its negative). The retarded response function has a finite imaginary part only in these regions and is purely real in the other intervals of the real frequency axis. 

The quantity $\chi_{\rm u}^{(0)}\left(\bm{q}, z\right)$ is singular for $r^2>0$, which corresponds to the kinematic threshold for excitation of pairs of momentum $q$. The variable $r$ is not well defined, because Eq.~(\ref{eq_definition_r}) presents a branch cut. The ambiguity in the sign of the square root is resolved by the $i\eta$ prescription. In the case of the retarded response function, $r$ is a positive quantity for $\omega$ real and greater than the threshold set by $r^{2}> 0$. In presence of a finite density $n$ of conduction electrons, low-energy inter-band transitions are Pauli blocked. Therefore, the inter-band contribution $\chi_{\rm inter}^{(0)}(\bm{q},z)$ to the density-density response function presents a finite imaginary part and branch-cut singularities only when $r^{2}> 0$ and $\left(k_{0}+r\right)>k_{\rm F}$. The inter-band electron-hole continuum in the $(q,\omega)$ plane is illustrated in Fig.~\ref{fig:figure2} for the case $m_{\rm c}=m_{\rm v}$.  

Analytical expressions for ${\rm Im}[\chi_{\rm u}^{(0)}(\bm{q},\omega)]$ and ${\rm Im}[\delta \chi_{\rm inter}^{(0)}(\bm{q}, \omega)]$ on the real frequency axis are reported in Appendix~\ref{appendix:appendix1}, see~Eqs.~(\ref{eq_undoped_imaginary_part}) and~(\ref{eq_delta_imaginary_part_antisymmetrization}). 

Results for the inter-band contribution
\begin{equation}\label{eq:interband_dielectric}
\epsilon_{\rm inter}(\bm{q},\omega)  \equiv  - \epsilon_{\rm b} v_{\bm{q}}
\chi^{(0)}_{\rm inter}(\bm{q},\omega+i0^{+})
\end{equation}
to the RPA dielectric function (\ref{eq_dielectric_function_RPA}) are shown in Fig.~\ref{fig:figure3} for undoped and doped LSMs. Note that rescaling $q$ with $k_{\rm F}$, $\omega$ with $E_{\rm F}/\hbar$, and $\epsilon_{\rm inter}(\bm{q},\omega)$ with $\epsilon_{\rm b}/(k_{\rm F}a_{\rm B})$, the plots shown in Fig.~\ref{fig:figure3} are universal, in the sense that they are independent of density, although they depend on the mass asymmetry parameter $\alpha$.

In the static  $z \to 0$ limit, $\chi^{(0)}_{\rm inter}$ reduces, in agreement with Ref.~\onlinecite{bardyszewski_jpcs_1983}, to
\begin{equation} \label{eq:chiinter_static}
\begin{split}
\chi_{\rm inter}^{(0)}(\bm{q}, 0)  &= -\frac{3}{4} N\left(E_{\rm F}\right) \bar{q}\Bigg\{\frac{\alpha}{2 \bar{q}} +  \frac{\bar{q}^{4} -1}{4\bar{q}^{2}} \ln \left|\frac{1 - \bar{q}}{1 + \bar{q}}\right|\\ & + \frac{(\alpha + 1)^{2} - \bar{q}^{4}}{8 \bar{q}^{2}} \ln \left[\frac{(\bar{q} - 1)^{2} + \alpha }{(\bar{q} + 1)^{2} + \alpha}\right]- \frac{\pi}{2}\sqrt{\alpha} \\  & + \sqrt{\alpha}\arctan \left(\frac{\alpha + 1 - \bar{q}^{2}}{2 \sqrt{\alpha}\,\bar{q}}\right)  + \Phi\left(\bar{q}\right) - \Phi\left(-\bar{q}\right) \\  + \left(\alpha - 1\right)& \mathrm{Re}\left[\mathrm{Li}_{2}\left(\frac{\bar{q}}{1 - i \sqrt{\alpha}}\right) - \mathrm{Li}_{2}\left(\frac{-\bar{q}}{1 - i \sqrt{\alpha}}\right) \right]\Bigg\}~. 
\end{split}
\end{equation}

\section{Plasmon dispersion relation and Landau damping}
\label{sect:plasmons}

In this section we use the results on the density-density response function to look at plasmons of doped LSMs.

The dispersion relation of plasmons is determined by the poles of the density response function, or equivalently, by the zeros of the dielectric function
\begin{equation}
\epsilon(\bm{q},\Omega(\bm{q})) = 0
\end{equation}
located infinitesimally below the real-frequency axis.

We begin by addressing the evaluation of the plasma frequency, i.e.~the value of $\Omega(\bm{q})$ at $\bm{q} = 0$, which we denote by $\Omega_{0}$. The real and imaginary parts of the RPA dielectric function in the $\bm{q}\to 0$ limit (and $\omega>0$) have the form:
\begin{eqnarray}
\label{eq_dynamical_dielectric_function_plasmons}
{\rm Re}[\epsilon(0,\omega)]  &=& \epsilon_{\rm b}
\left\{1 - \frac{\omega_{\rm p}^{2}}{\omega^{2}} +\frac{3}{4}\frac{\alpha}{\alpha+1}
\frac{\hbar^{2}\omega^2_{\rm p}}{E_{\rm F}^{2}}\right.
\nonumber\\ 
& \times&
\left.\frac{1}{\bar{r}}
\left[
\frac{1}{4}\ln\frac{(1+\bar{r})^{2}}{(1-\bar{r})^{2}}+\arctan(\bar{r})
\right]
\right\}
\end{eqnarray}
and
\begin{equation}
\label{eq_dynamical_dielectric_function_imag}
{\rm Im}[\epsilon(0,\omega)]  = \frac{3\pi}{8} \epsilon_{\rm b}\frac{\alpha }{1+\alpha}  \frac{\hbar^{2} \omega^{2}_{\rm p}}{E_{\rm F}^{2}} \frac{\theta(\bar{r} - 1)}{\bar{r}}~,
\end{equation}
where $\omega^2_{\rm p} = 4 \pi n e^{2}/(\epsilon_{\rm b} m_{\rm c})$ is the square of the standard plasma frequency for a 3D systems of electrons with a single parabolic band of mass $m_{\rm c}$~\cite{giuliani_and_vignale}, $\bar{r}$ is the variable introduced previously in Eq.~(\ref{eq_definition_r}), evaluated at zero momentum, i.e.~,
\begin{equation}
\bar{r} = \frac{1}{k_{\rm F}} \sqrt{\frac{\alpha}{\alpha+1} \frac{2 m_{\rm c} \omega}{\hbar}}~,
\end{equation}
and $\theta(x)$ is the usual step function.
These expressions were first obtained in Ref.~\onlinecite{broerman_prb_1972}. The three terms in the right-hand side of Eq.~(\ref{eq_dynamical_dielectric_function_plasmons}) correspond to the background, intra-band, and inter-band contributions to the $q=0$ frequency-dependent dielectric function, respectively. Only inter-band transitions contribute to the imaginary part, Eq.~(\ref{eq_dynamical_dielectric_function_imag}). These lead to singularities in the dielectric function~\cite{sherrington_prl_1968, broerman_prb_1972, cheng_ncomm2017} for $\bar{r} = 1$. The imaginary part presents a finite jump and the real part a logarithmic divergence~\cite{sherrington_prl_1968, broerman_prb_1972, cheng_ncomm2017}. Such singularities disappear at finite temperature~\cite{broerman_prb_1972}. The frequency dependence of ${\rm Re}[\epsilon(0, \omega)]$ and ${\rm Im}[\epsilon(q, \omega)]$ is shown in Fig.~\ref{fig:figure3}(a). 

In the limit $r_s \ll 1$, the plasmon frequency is given by~\cite{goettig_pssb_1975}
\begin{equation}
\label{eq_plasmon_frequency_weak_coupling}
\Omega^2_{0} = \frac{\omega_{\rm p}^{2}}{1+\epsilon_{\rm inter}(0,0)/\epsilon_{\rm b}}~,
\end{equation}
where
\begin{equation}\label{eq:qzeroomegazero}
\epsilon_{\rm inter}(0,0) = \frac{3}{2}\epsilon_{\rm b}\frac{\alpha}{\alpha+1} \frac{\hbar^{2}\omega_{\rm p}^{2}}{E_{\rm F}^{2}}~.
\end{equation} 
In the limit $r_s\ll 1$, indeed, the inter-band contribution is small, because $\hbar^{2}\omega_{\rm p}^{2}/E_{\rm F}^{2} \approx r_s$ and $\bar{r}\approx \sqrt{\hbar \omega_{\rm p}/E_{\rm F}} \approx r^{1/4}_s$, and gives a contribution of order $r_s$.
When the electron density is lowered, the frequency dependence of the dynamical inter-band dielectric function plays a role and the plasma frequency is significantly modified~\cite{goettig_pssb_1975} with respect to Eqs.~(\ref{eq_plasmon_frequency_weak_coupling})-(\ref{eq:qzeroomegazero}). We remind the reader that, however, the RPA is not reliable in the low-density limit~\cite{giuliani_and_vignale} and we will therefore mainly stick to discussing the weak-coupling $r_s\lesssim 1$ regime.

\begin{figure}[h]
\centering
\begin{overpic}[width=\columnwidth]{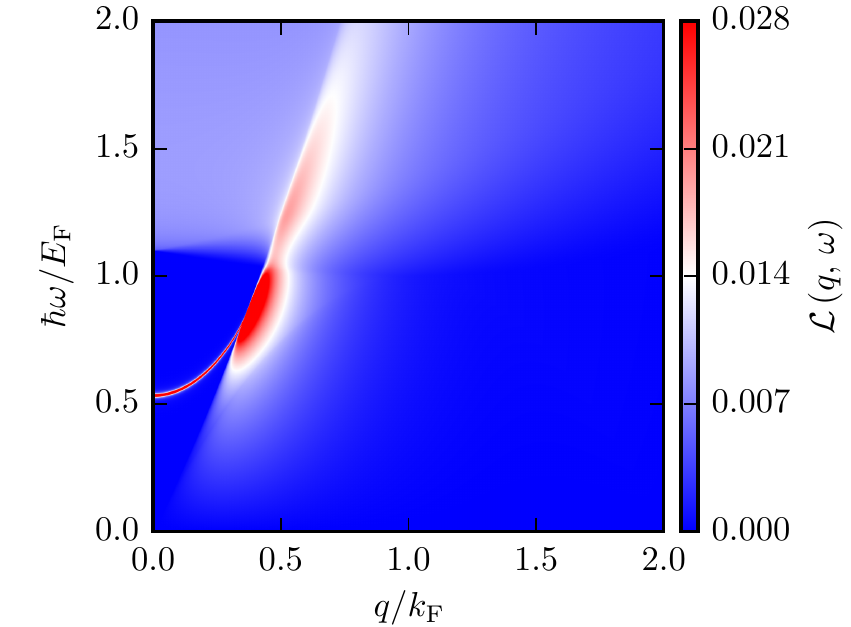}
\put(1,75){(a)}
\end{overpic}
\begin{overpic}[width=\columnwidth]{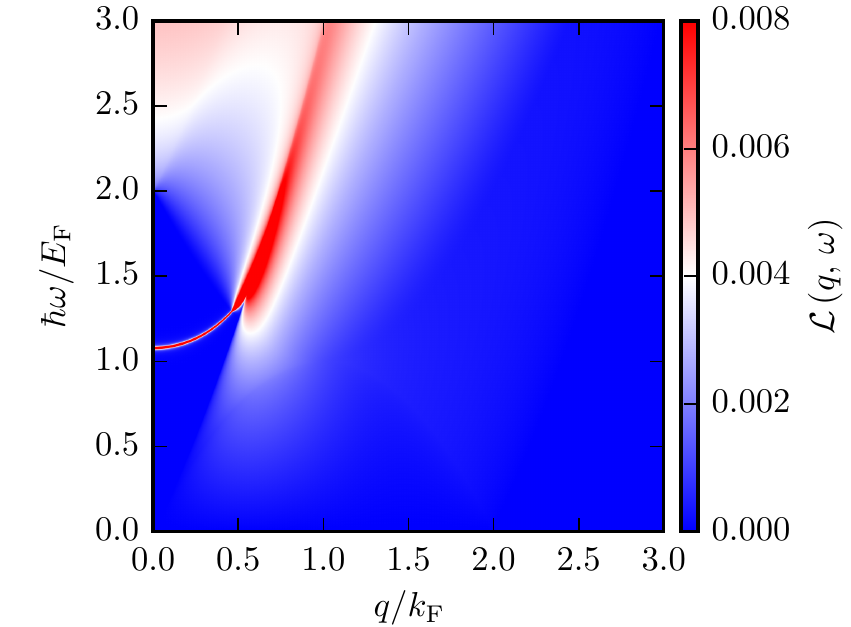}
\put(1,75){(b)}
\end{overpic}
\caption{\label{fig:figure4}
Density plots of the loss function for a doped Luttinger semimetal. 
(a) ${\cal L}\left(\bm{q}, \omega\right)$ for $m_{\rm c} = 0.024~m_{\rm e}$, $m_{\rm v} = 10~m_{\rm v}$, $\epsilon_{\rm b} = 24$, and $n = 10^{16}$ cm$^{-3}$ (corresponding approximately to material parameters for $\alpha$-Sn). (b) Same as in (a) but for $m_{\rm c} = m_{\rm v} = 6.3~m_{\rm e}$, $\epsilon_{\rm b} = 10$, and $n = 5.1 \times 10^{19}~$cm$^{-3}$ (corresponding approximately to material parameters for Pr$_{2}$Ir$_{2}$O$_{7}$, see Refs.~\onlinecite{kondo_ncomm_2015,cheng_ncomm2017}). Results have been obtained by giving a finite imaginary shift to the frequency, which artificially broadens the plasmon dispersion relation, mimicking the role of extrinsic disorder. In both figures, the color intensity is saturated when the loss function exceeds the maximum in the scale.}
\end{figure}

In order to address the plasmon dispersion relation at finite momenta, transcending the analytical result (\ref{eq_plasmon_frequency_weak_coupling}), we seek for plasmons numerically by plotting the loss function
\begin{equation}
{\cal L}(\bm{q},\omega) = -{\rm Im}\left[\frac{1}{\epsilon(\bm{q},\omega+i0^+)}\right]~,
\end{equation}
which portraits the spectral density of particle-hole excitations and presents sharp peaks in correspondence of plasmons. Illustrative numerical results for ${\cal L}(\bm{q},\omega)$, which can be measured experimentally by electron-energy loss spectroscopy (EELS)~\cite{eels}, are reported in Figs.~\ref{fig:figure4}(a) and~(b). 

Data in Fig.~\ref{fig:figure4}(a) have been obtained by setting $m_{\rm c}=0.024~m_{\rm e}$, $m_{\rm v}= 10~m_{\rm c}$, $\epsilon_{\rm b}=24$, and a conduction electron density $n=10^{16}~{\rm cm}^{-3}$---corresponding to a value of the Wigner-Seitz coupling constant $r_{\rm s}\simeq 0.55$. Here, $m_{\rm e}$ is the bare electron mass in vacuum. These parameters are supposed to describe $\alpha$-Sn. Material parameters for the pyrochlore iridate Pr$_{2}$Ir$_{2}$O$_{7}$ are very different: we take~\cite{kondo_ncomm_2015,cheng_ncomm2017} $m_{\rm c} = m_{\rm v} = 6.3~m_{\rm e}$ and $\epsilon_{\rm b} = 10$. A finite carrier concentration is generally expected in a real sample, due to the unavoidable presence of impurities. Samples of Pr$_{2}$Ir$_{2}$O$_{7}$ prepared in Ref.~\onlinecite{cheng_ncomm2017} were indeed found to present a hole carrier density in the range $5.1 \times 10^{19}~{\rm cm}^{-3}$- $1.7 \times 10^{20}~{\rm cm}^{-3}$.  In this case, the value of the coupling constant $r_s$ is much larger than the one for $\alpha$-Sn: in Pr$_{2}$Ir$_{2}$O$_{7}$, $r_{s}$ is expected to vary  between approximately 10 and 20. In Fig.~\ref{fig:figure4}(b), we report the loss function ${\cal L}(\bm{q}, \omega)$ corresponding to these material parameters and for a hole density $n = 5.1 \times 10^{19}$ cm$^{-3}$ (corresponding to $r_{\rm s}\simeq 20$). Due to the large effective masses, interaction effects are enhanced in Pr$_{2}$Ir$_{2}$O$_{7}$, which makes this material a specially suitable candidate for the experimental observation of non-Fermi liquid behavior predicted by Abrikosov~\cite{abrikosov_jetp_1971, abrikosov_jetp_1974, abrikosov_jltp_1971, abrikosov_jltp_1975, moon_prl_2013, cheng_ncomm2017, kondo_ncomm_2015} or broken-symmetry phases induced by electron-electron interactions~\cite{janssen_prb_2015, herbut_prl_2014, boettcher_prb_2017, janssen_prb_2016, janssen_prb_2017, goswami_prb_2017}. In Ref.~\onlinecite{cheng_ncomm2017}, it is argued that dc transport measurements and Terahertz spectroscopy on samples of Pr$_{2}$Ir$_{2}$O$_{7}$ with a carrier density larger than $5.1 \times 10^{19}$~cm$^{-3}$ can be consistently understood assuming a Fermi-liquid behaviour at low energy and low temperatures. At the same time, experimental evidence of anomalously strong dielectric response was reported~\cite{cheng_ncomm2017}. This was interpreted as a result of a dielectric anomaly due to inter-band transitions at the quadratic band touching, which according to Eq.~\ref{eq_dynamical_dielectric_function_plasmons} and~\ref{eq:qzeroomegazero}, is a signature of strong-coupling and large values of $r_{\rm s}$. For coupling constants as large as  $r_{\rm s} = 10$--$20$, the applicability of the RPA is not justified and results obtained in our work are not expected to be quantitatively accurate. In Ref.~\onlinecite{cheng_ncomm2017}, it was estimated, based on the theory of Abrikosov~\cite{abrikosov_jetp_1971, abrikosov_jetp_1974, abrikosov_jltp_1971, abrikosov_jltp_1975}, that, despite the large value of the coupling, the dielectric anomaly observed in Pr$_{2}$Ir$_{2}$O$_{7}$ is accurately captured by the RPA in the range of experimental parameters. In this work, however, we will mainly focus on the weak-coupling regime $r_{\rm s} \lesssim 1$.

\begin{figure}[t]
\centering
\begin{overpic}[width=\columnwidth]{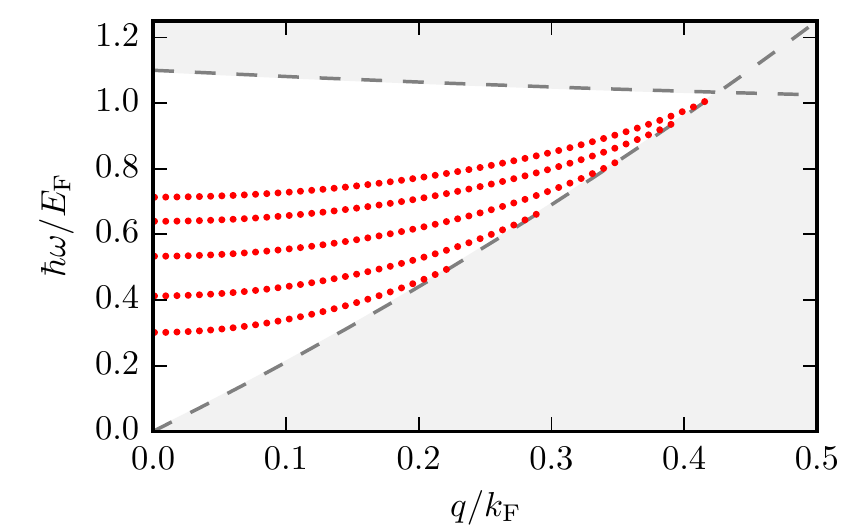}
\end{overpic}
\caption{\label{fig:figure5} Plasmon dispersion relation (red dots), calculated for $m_{\rm c} = 0.024~m_{\rm e}$, $m_{\rm v} = 10~m_{\rm e}$, $\epsilon_{\rm b} = 24$, and for different conduction electron densities: from the bottom to the top, $n = 10^{18}$, $10^{17}$, $10^{16}$, $10^{15}$, and $10^{14}~{\rm cm}^{-3}$.  Grey-shaded regions denote the intra- and inter-band particle-hole continua. In these regions the imaginary part of the non-interacting density response function is non-vanishing and plasmons are Landau damped: see Fig.~\ref{fig:figure4}(a).}
\end{figure}
Numerical results for the plasmon dispersion are shown in Fig.~\ref{fig:figure5}, for the following choice of parameters: $m_{\rm c} = 0.024~m_{\rm e}$, $m_{\rm v}= 10~m_{\rm c}$, and $\epsilon_{\rm b}=24$. The dispersion relation $\Omega=\Omega(\bm{q})$  is shown for values of the electron density  between $10^{18}~{\rm cm}^{-3}$ and $10^{14}~{\rm cm}^{-3}$, which correspond to values of $r_{s}$ between $\simeq 0.12$ and $\simeq 2.5$. In the RPA, the plasmon lives in region III of Fig.~\ref{fig:figure2}, where the spectral density of particle-hole pairs is zero, until a finite critical wave vector, after which it merges with the particle-hole continuum and ceases to exist as a sharply defined excitation because Landau damping kicks in. Now, this occurs~\cite{goettig_pssb_1975} for arbitrarily low values of $n$, for a finite range of wave vectors $q$, which shrinks in size as $n\to 0$. Indeed, for arbitrary values of $n$, the dynamic dielectric function at $\bm{q}=0$---Eq.~(\ref{eq_dynamical_dielectric_function_plasmons})---diverges to $-\infty$ for $\omega \rightarrow 0$ due to the intra-band contribution and to $+\infty$ for $\bar{r}\rightarrow 1$ due to the inter-band contribution. Therefore, the zero of $\epsilon(0,\omega)$, which yields the plasmon pole $\Omega(0)$ at $\bm{q}=0$, must occur for $\bar{r}<1$, where $\epsilon$ is purely real and Landau damping does not occur. By continuity,  if the plasmon mode lies outside the phase-space boundaries of single particle-hole excitations at $\bm{q}=0$, its dispersion must persist in region III of Fig.~\ref{fig:figure2} for a finite range of wave vectors~\cite{goettig_pssb_1975}. 

Such a simple picture is valid as long as beyond-RPA effects are neglected. In reality, intrinsic plasmon damping due to the excitation of double particle-hole pairs with opposite momenta persists down to $\bm{q}=0$ and can be calculated by diagrammatic perturbation theory applied to the proper density-density response function~\cite{giuliani_and_vignale,principi_prb_2013}.

\section{Summary and conclusions}
\label{sect:summary}

In summary, in this Article we have presented analytical formulas for the non-interacting density-density response function $\chi^{(0)}_{nn}(\bm{q},z)$ of a doped Luttinger semimetal. Our results are valid for arbitrary values of the wave vector $\bm{q}$, complex frequency $z$, and conduction-valence effective mass asymmetry $\alpha$. 

The intra-band contribution can be found in Eq.~(\ref{eq_intra}). The inter-band contribution can be found in Eqs.~(\ref{eq_undoped_interband_response_function})-(\ref{eq_doped_interband_response_function}). These results have been employed as ingredients for the calculation of the interacting density-density response $\chi_{nn}(\bm{q},\omega)$ and dynamical dielectric function $\epsilon(\bm{q},\omega)$ in the celebrated random phase approximation~\cite{giuliani_and_vignale}, Eqs.~(\ref{eq_random_phase_approximation}) and~(\ref{eq_dielectric_function_RPA}), respectively. Plasmons at the level of RPA have been discussed in Sect.~\ref{sect:plasmons}.

The range of applicability of these results is however much wider. For example, $\chi^{(0)}_{nn}(\bm{q},z)$ is the key ingredient of any calculation of normal Fermi liquid properties~\cite{giuliani_and_vignale}, with particular reference to the spectral function $A({\bm k},\omega)$---see Ref.~\onlinecite{tchoumakov_arxiv_2019} for calculations at $\alpha=1$---and Landau parameters $F^{({\rm s}, {\rm a})}_{\ell}$. Indeed, these calculations usually rely on the dynamically-screened electron-electron interaction
\begin{equation}\label{eq:screened_interaction}
W(\bm{q},z) \equiv \frac{v_{\bm{q}}}{\epsilon(\bm{q},z)}~.
\end{equation}
For example, all calculations of the quasiparticle self-energy $\Sigma({\bm k},\omega)$ based on the so-called ``GW" approximation rely on Eq.~(\ref{eq:screened_interaction}).
In general, any diagrammatic calculation aimed at transcending the RPA---by relaxing the brutal approximation of the proper density-density response function $\tilde{\chi}_{nn}(\bm{q},z)$ with $\chi^{(0)}_{nn}(\bm{q},z)$---requires the analytical formulas reported in this Article.

\acknowledgments
This work was supported by the European Union's Horizon 2020 research and innovation programme under grant agreement No. 785219 - GrapheneCore2.
We wish to thank A. Tomadin, M. Beconcini, and G.M. Andolina for useful discussions.

\appendix

\counterwithin{figure}{section}
\section{Calculation of the inter-band contribution} 
\label{appendix:appendix1}

The calculation of the inter-band contribution to the density-density response function starts from the usual Kubo formula, Eq.~\ref{eq_kubo_formula_interband_contributions}. As in the case of the intra-band term, we first calculate the imaginary part of the retarded density-density response function $\chi_{\rm R, inter}^{(0)}(\bm{q},\omega) = \chi_{\rm inter}^{(0)}(\bm{q},\omega + i 0^{+})$. We then retrieve the complete analytic expression for $\chi^{(0)}_{\rm inter}(\bm{q},z)$ as an analytic function in the complex frequency plane from the Kramers-Kronig relation~\cite{giuliani_and_vignale}. 

For the sake of simplicity, in this Appendix we set $\hbar = 2m_{\rm c} = k_{\rm F} = 1$. Final results are presented in ordinary units in the main text. We remind the reader of the definition of the effective mass imbalance $\alpha = m_{\rm v}/m_{\rm c}$. 

The undoped response function is obtained by setting $n_{\bm{k}, {\rm c}} = 0$ in Eq.~(\ref{eq_kubo_formula_interband_contributions}):
\begin{widetext}
\begin{equation}\label{eq_undoped_imaginary_part_kubo_formula}
{\rm Im}[\chi_{\rm R, u}^{(0)}(\bm{q},\omega)] \equiv {\rm Im}[\chi_{\rm u}^{(0)}(\bm{q},\omega + i 0^{+})] = - \pi \int \frac{{\rm d}^{3}\bm{k}}{\left(2 \pi\right)^{3}} \frac{3}{2} \frac{\left(\bm{q}\times \bm{k}\right)^{2}}{k^{2}\left(\bm{k}-\bm{q}\right)^{2}}\delta \left( \omega -\epsilon_{\bm{k},{\rm c}}+ \epsilon_{\bm{k}-\bm{q},{\rm v}}\right) - [\omega \rightarrow -\omega]
\end{equation}
\end{widetext}
${\rm Im}[\chi_{\rm R, u}^{(0)}(\bm{q},\omega)]$ is non-zero only in the region of the $(q,\omega)$ plane where inter-band particle-hole excitations are possible according to the conservation of energy and momentum. This region corresponds to $\omega > q^{2}/(1+\alpha)$. (We can restrict our attention to the case of positive $\omega$, because of the antisymmetry of the imaginary part of the response function.)

When this condition is fulfilled, the integral in Eq.~(\ref{eq_undoped_imaginary_part_kubo_formula}), restricted by the energy conserving $\delta$ function, must be carried out over the locus of all final momenta $\bm{k}$ available to electrons being promoted from the valence to the conduction band with an energy transfer $ \omega$ and a momentum transfer $\bm{q}$. This is a sphere, whose center is located at
\begin{equation}
\label{eq_center_interband_transitions}
\bm{k}_{0} =\frac{\bm{q}}{\alpha+1}
\end{equation}
and whose radius is
\begin{equation}
\label{eq_radius_interband_transitions}
\rho = \sqrt{ \frac{\alpha}{\alpha+1} \left(\omega-\frac{q^{2}}{\alpha+1}\right)}~.
\end{equation}
The integral in Eq.~(\ref{eq_undoped_imaginary_part_kubo_formula}) can be calculated analytically and conveniently expressed in terms of the variables $\rho$ and $k_{0}$:
\begin{widetext}
\begin{eqnarray}
\label{eq_undoped_imaginary_part}
{\rm Im}[\chi_{\rm R, u}^{(0)}(\bm{q},\omega)] &=&  -\frac{3}{64\pi}\Theta\left(\rho^{2}\right)\left[2\left(\alpha + 1\right)\rho + \frac{1}{2\alpha k_{0}}\frac{\left(\rho^2-\alpha^2k_{0}^2\right)^2}{\rho^2+\alpha k_{0}^2}\ln\frac{\left(\rho-\alpha k_{0}\right)^2}{\left(\rho+\alpha k_{0}\right)^2}-\frac{\alpha}{2k_{0}}\frac{\left(\rho^2-k_{0}^2\right)^2}{\rho^2+\alpha k_0^2}\ln\frac{\left(\rho+k_0\right)^2}{\left(\rho-k_0\right)^2}\right]\nonumber\\
&-& \left[\omega \rightarrow -\omega\right]~.
\end{eqnarray}
\end{widetext}
In the following we will make use of the variable
\begin{equation}
\label{eq_analytic_radius_interband_transitions}
r = i\sqrt{\frac{\alpha}{\alpha+1} \left(\frac{q^{2}}{\alpha+1}- z\right)}~,
\end{equation}
which is very similar to $\rho$, but for the presence of a branch cut in the expected region, i.e.~for real frequencies $z$ such that $z > q^{2}/(1+\alpha)$. In Eq.~(\ref{eq_analytic_radius_interband_transitions}), as in the main text, the square root is intended to have a branch cut for negative values of its argument and a positive real part. In the following calculations, advantages in using $r$ arise from the fact that, for every complex value of $z$ (apart from values which lie on the branch cut on the real axis, which are never reached thanks to the $+i \eta$ prescription), the imaginary part of $r$ is strictly positive. 

We now calculate the density response function $\chi^{(0)}_{\rm u}(\bm{q},z)$ from the Kramers-Kronig relation. Changing variables from $\omega^{\prime}$ to $\rho^{\prime}$ and making use of the antisymmetry of the expression in the first line of Eq.~(\ref{eq_undoped_imaginary_part}) under the exchange $\rho \rightarrow -\rho$ yields:
\begin{widetext}
\begin{equation}
\label{eq_undoped_response_function}
\begin{split}
&\chi_{\rm u}^{(0)}(\bm{q},z) = \frac{1}{\pi}\int_{-\infty}^{\infty}{\rm d}\omega^{\prime} \frac{{\rm Im}[\chi_{\rm R, u}^{(0)}(\bm{q},\omega^{\prime})]}{\omega^{\prime}-z} 
\\ = & -\frac{3}{64\pi^{2}}\int_{-\infty}^{\infty} \frac{{\rm d}\rho^{\prime}}{\rho^{\prime}-r} \left[2\left(\alpha + 1\right)\rho^{\prime} + \frac{1}{2\alpha k_{0}}\frac{\left(\rho^{\prime 2}-\alpha^2k_{0}^2\right)^2}{\rho^{\prime 2}+\alpha k_{0}^2}\ln\frac{\left(\rho^{\prime}-\alpha k_{0}\right)^2}{\left(\rho^{\prime}+\alpha k_{0}\right)^2}-\frac{\alpha}{2k_{0}}\frac{\left(\rho^{\prime 2}-k_{0}^2\right)^2}{\rho^{\prime 2}+\alpha k_0^2}\ln\frac{\left(\rho^{\prime}+k_0\right)^2}{\left(\rho^{\prime}-k_0\right)^2}\right] \nonumber\\
&+ \left[z \rightarrow -z\right]~.
\end{split}
\end{equation}
The integral in Eq.~(\ref{eq_undoped_response_function}) can be calculated with the aid of the relation
\begin{equation}
\int_{-\infty}^{+\infty}\frac{{\rm d}x}{ x-z} \ln \frac{\left(x+1\right)^{2}}{\left(x-1\right)^{2}} = 2\pi i {\rm sgn}[{\rm Im}(z)]\ln\left(\frac{z+1}{z-1}\right) = 4\pi {\rm sgn}\left[{\rm Im}(z)\right] \arctan\left(\frac{i}{z}\right)~.
\end{equation}
The final result is:
\begin{eqnarray}
\chi_{\rm u}^{(0)}(\bm{q},z) = -\frac{3 i}{32\pi}\left(\alpha+1\right)r &+&  \frac{3}{32\pi k_0}\frac{\alpha}{r^2+\alpha k_0^2}\left[-\frac{\pi}{2} k_{0}^{4}\left(\alpha+1\right)^2+\left(r^2-k_{0}^2\right)^2\arctan\left(\frac{ik_{0}}{r}\right)\right.\nonumber\\
&+&\left.\alpha^2\left(\frac{r^2}{\alpha^2}-k_{0}^2\right)^2\arctan\left(\frac{i\alpha k_0}{r}\right)\right]+ {\left[z\rightarrow-z\right]}~.
\end{eqnarray}
Symmetrizing with respect to $z \propto r^{2}+\alpha k_{0}^{2}$ we find
\begin{equation}
\label{eq_undoped_response_function_result}
\begin{split}
\chi_{\rm u}^{(0)}(\bm{q},z) &=  -\frac{3 i}{32\pi}\left(\alpha+1\right)r+  \frac{3}{32\pi k_0}\frac{\alpha}{r^2+\alpha k_0^2}\left[\left(r^2-k_{0}^2\right)^2\arctan\left(\frac{ik_{0}}{r}\right)+\alpha^2\left(\frac{r^2}{\alpha^2}-k_{0}^2\right)^2\arctan\left(\frac{i\alpha k_0}{r}\right)\right]\\ &+ {\left[z\rightarrow-z\right]}~,
\end{split}
\end{equation}
\end{widetext}
which, after using Eqs.~(\ref{eq_definition_r}) and~(\ref{eq_definition_k0}), yields Eq.~(\ref{eq_undoped_interband_response_function}) in the main text. 

The analytic expression in Eq.~(\ref{eq_undoped_response_function_result}) presents a single branch cut in the expected region ($r^{2}>0$) in which inter-band electron-hole excitations exist. The $\arctan$ functions produce branch cut singularities only in the same region, $r^{2}> 0$. Note that the density response function does not present a pole at $z \propto r^{2}+\alpha k_{0}^{2} = 0$ because it is symmetric under exchange $z\rightarrow -z$. The imaginary part of $\chi_{\rm R, u}^{(0)}(\bm{q},\omega)$---Eq.~(\ref{eq_undoped_imaginary_part})---can be obtained from Eq.~(\ref{eq_undoped_response_function_result}) by taking the limit $\lim_{\eta \rightarrow 0^+}\chi_{\rm u}^{(0)}(\bm{q},\omega+i \eta)$ and therefore does not need to be considered separately. 

We now turn to the contribution $\delta \chi_{\rm inter}^{(0)}(q,z)$, which arises from the presence of a finite density of conduction-band electrons. According to Eq.~(\ref{eq_kubo_formula_interband_contributions}) it is given by:
\begin{equation}
\begin{split}
\delta \chi_{\rm inter}^{(0)}(\bm{q},z) = & \frac{1}{V} \sum_{\bm{k}} \frac{3}{2} \sin^{2}(\theta_{\bm{k}-\bm{q}, \bm{k}})\frac{-n_{\bm{k}, {\rm c}}}{\hbar z +\epsilon_{\bm{k}-\bm{q}, {\rm v}} -\epsilon_{\bm{k}, {\rm c}}} \\ & + \left[z \rightarrow -z\right]~.\\
\end{split}
\end{equation}
The imaginary part on the real frequency axis is therefore given by
\begin{widetext}
\begin{equation}
 \label{eq_delta_imaginary_part_kubo_formula}
\begin{split}
{\rm Im}[\delta \chi_{\rm R, inter}^{(0)}(\bm{q},\omega)] \equiv {\rm Im}[\delta \chi_{\rm inter}^{(0)}(\bm{q},\omega + i 0^{+})] &= - \pi \int \frac{{\rm d}^{3}\bm{k}}{\left(2 \pi\right)^{3}} \frac{3}{2} \frac{\left(\bm{q}\times \bm{k}\right)^{2}}{k^{2}\left(\bm{k}-\bm{q}\right)^{2}}\left(-n_{\bm{k},{\rm c}}\right)\delta \left( \omega -\epsilon_{\bm{k},{\rm c}}+ \epsilon_{\bm{k}-\bm{q},{\rm v}}\right) \\
&- \left[\omega \rightarrow -\omega\right]~.\\ 
\end{split}
\end{equation}
\end{widetext}
We can again limit our attention to $\omega>0$, because of the antisymmetry of the previous response function. As in the calculation of the density-density response function in the undoped case, the integral over ${\bm k}$ (constrained by the energy conserving $\delta$ function) spans a sphere with center $\bm{k}_{0}$ and radius $\rho$. The quantity ${\rm Im}[\delta \chi_{\rm R,  inter}^{(0)}(\bm{q},\omega)]$ acts to subtract from ${\rm Im}[\chi_{\rm R,  u}^{(0)}(\bm{q},\omega)]$---Eq.~(\ref{eq_undoped_imaginary_part})---those transitions whose final momentum $\bm{k}$ would lie inside the occupied Fermi sphere. Depending on $q$ and $\omega$ (and therefore on $k_{0}$ and $\rho$) the Pauli principle can block all of the transitions, a part, or none of them. The integral in Eq.~(\ref{eq_delta_imaginary_part_kubo_formula}) can be calculated in cylindrical coordinates, with the axis oriented along ${\bm q}$. Contracting the $\delta$ function with the integration over the radial cylindrical coordinate $k_{\perp}$ yields: 
\begin{widetext}
\begin{equation}
\label{eq_delta_imaginary_part_antisymmetrization}
\begin{split}
{\rm Im}[\delta \chi_{\rm R,  inter}^{(0)}(\bm{q},\omega)] & = \frac{3}{16\pi} \frac{\alpha}{\alpha+1}\theta\left(\rho^{2}\right)\theta\left(1-\left(k_{0}-\rho\right)^{2}\right)\int_{k_{0}-\rho}^{\min \left(k_{0}+\rho, \frac{1-\rho^{2}+k_{0}^{2}}{2k_{0}}\right)} {\rm d}k_{\parallel} \frac{q^{2} k_{\perp}^{2}}{(k_{\parallel}^{2}+k_{\perp}^{2})\left[k_{\perp}^{2}+(k_{\parallel}-q)^{2}\right]}\\ & - \left[\omega\rightarrow -\omega\right]~,
\end{split}
\end{equation}
where $k_{\perp}^{2} = \rho^{2}-\left(k_{\parallel}-k_{0}\right)^{2}$. The final result is:
\begin{equation}
\label{eq_delta_imaginary_part}
\begin{split}
\mathrm{Im}[\delta \chi_{\rm R, inter}^{(0)}(\bm{q}, \omega)] =& \frac{3}{64\pi} \theta\left(\rho^{2}\right) \Bigg\{\theta\left(1-\left(\rho-k_{0}\right)^{2}\right)\Bigg[\left(\alpha+1\right)\frac{1-\left(\rho-k_0\right)^2}{2k_0}+\frac{\alpha}{2k_0}\frac{\left(\rho^2-k_0^2\right)^2}{\rho^2+\alpha k_0^2}\ln\left(\rho-k_0\right)^{2}\\ & -\frac{1}{2\alpha k_0}\frac{\left(\rho^2-\alpha^2k_0^2\right)^2}{\rho^2+\alpha k_{0}^2} \ln \left(\rho+\alpha k_0\right)^2\Bigg] - \theta\left(1-\left(\rho+k_{0}\right)^{2}\right)\Big[\dots\Big]_{\rho\rightarrow -\rho}\\ &
+\left[\theta\left(1-\left(\rho-k_{0}\right)^{2}\right)-\theta\left(1-\left(\rho+k_{0}\right)^{2}\right)\right] \frac{1}{2\alpha k_0}\frac{\left(\rho^2-\alpha^2k_0^2\right)^2}{\rho^2+\alpha k_{0}^2} \ln \left[\left(\alpha+1\right)\left(\rho^2+\alpha k_0^2\right)-\alpha\right] \Bigg\}
\\ & - \left[\omega \rightarrow -\omega\right]~.
\end{split}
\end{equation}
\end{widetext}
Here, the symbol ``$[...]_{\rho \rightarrow -\rho}$" denotes the expression obtained by changing the sign of $\rho$ in the terms in the squared bracket in the first and second lines of Eq.~(\ref{eq_delta_imaginary_part}). As in the rest of the paper, $[\omega \rightarrow -\omega]$ denotes the function obtained by changing the sign of $\omega$ in all terms in the first three lines. The same notation will be used in Eq.~(\ref{eq_delta_kramers_kronig}). The boundaries set by the Heaviside step functions in Eq.~(\ref{eq_delta_imaginary_part}) have a transparent physical interpretation. The electron momenta of interband electron-hole excitations of energy $\omega$ and momentum $q$ lie on a spherical surface. If $-1<k_{0}-\rho<1$, this surface intersects the Fermi sphere and therefore some of the transitions will be denied by the Pauli principle. If furthermore $k_{0}+\rho < 1$, the surface lies entirely inside the Fermi sphere and all of the transitions are Pauli blocked. In this case, $\mathrm{Im}[\delta \chi_{\rm R, inter}^{(0)}(\bm{q},\omega)]$ becomes equal and opposite to $\mathrm{Im}[ \chi_{\rm R, u}^{(0)}(\bm{q},\omega)]$, as can be seen by comparing Eq.~(\ref{eq_delta_imaginary_part}) and Eq.~(\ref{eq_undoped_imaginary_part}).

 It remains to use the Kramers-Kronig relation to calculate $\delta\chi_{\rm inter}^{(0)}(q,z)$ from the knowledge of its imaginary part on the real frequency axis. As in the case of $\chi_{\rm u}(\bm{q},z)$, the calculation is simplified by the antisymmetry of the expressions defined by the first three lines of Eq.~(\ref{eq_delta_imaginary_part}) under the exchange $\rho \rightarrow -\rho$. We have
\begin{widetext}
\begin{equation}
\label{eq_delta_kramers_kronig}
\begin{split}
\delta \chi_{\rm inter}^{(0)}(\bm{q},z) &= \frac{1}{\pi}\int_{-\infty}^{\infty} {\rm d}\omega^{\prime} \frac{{\rm Im} \delta \chi_{\rm R, inter}^{(0)}(\bm{q},\omega^{\prime})}{\omega^{\prime}-z}\\ & = 
\Bigg\{\frac{3}{64\pi^2} \int _{-1+k_0}^{1+k_0}\frac{{\rm d}\rho^{\prime}}{\rho^{\prime}-r}
 \Big[\left(\alpha+1\right)\frac{1-\left(\rho^{\prime}-k_0\right)^2}{2k_0}+\frac{\alpha}{2k_0}\frac{\left(\rho^{\prime 2}-k_0^2\right)^2}{\rho^{\prime 2}+\alpha k_0^2}\ln\left(\rho^{\prime}-k_0\right)^{2} \\ &  -\frac{1}{2\alpha k_0}\frac{\left(\rho^{\prime 2}-\alpha^2k_0^2\right)^2}{\rho^{\prime 2}+\alpha k_{0}^2}\ln\left(\rho^{\prime}+\alpha k_0\right)^{2}\Big] \Bigg\}  + \Big\{\dots\Big\}_{r\rightarrow -r} \\ & + \frac{3}{64 \pi^{2}}\int_{\left(1-k_{0}\right)^{2}}^{\left(1+k_{0}\right)^{2}}\frac{{\rm d} \left(\rho^{\prime 2}\right)}{\rho^{\prime 2}-r^{2}} \frac{1}{2\alpha k_{0}}\frac{\left(\rho^{\prime 2}-\alpha^{2} k_{0}^{2}\right)^{2}}{\rho^{\prime 2}+\alpha k_{0}^{2}}\ln \left[(\alpha+1)(\rho^{\prime 2}+\alpha k_{0}^{2})-\alpha\right] \\
 &+ \left[z \rightarrow -z\right]~.
\end{split}
\end{equation}
The final result is:
\begin{equation}
\label{eq_delta_response_function_result}
\begin{split}
\delta\chi_{\rm inter}^{(0)}(\bm{q}, z)&  =  \frac{3}{64\pi^2}\Bigg\{ -\frac{1}{2k_0}\left[4\alpha k_0+\frac{1-(\alpha + 1)^2 k_0^2}{\alpha + 1}\ln \frac{\left(1+(\alpha + 1)k_0\right)^2}{\left(1-(\alpha + 1)k_0\right)^2}\right] +\frac{\alpha+1}{2k_0}\Bigg[\left((r-k_0)^2-1\right)\ln\frac{r-k_0+1}{r-k_0-1}\\ & -\left((r+k_0)^2-1\right)\ln\frac{r+k_0+1}{r+k_0-1}\Bigg] -\frac{1}{2\alpha k_0}\frac{\left(r^2-\alpha^2 k_0^2\right)^2}{r^2+\alpha k_0^2}\Bigg[f\left(-(\alpha+1)k_0,r-k_0\right)\\ & +f(-(\alpha+1)k_0,-r-k_0)-g\left(-(\alpha+1)k_0,\frac{r^2-1-k_0^2}{2k_0}\right)\Bigg] +\frac{\alpha}{2k_0}\frac{\left(r^2-k_0^2\right)^2}{r^2+\alpha k_0^2}[f(0,r-k_0)\\ & +f(0,-r-k_0)] \Bigg\}\\
&+\left[z \rightarrow -z\right]~,
\end{split}
\end{equation}
\end{widetext}
where the functions $f(\bar{q},t)$ and $g(\bar{q},t)$ have been defined in Eqs.~(\ref{eq_f}) and~(\ref{eq_g}). Combining Eq.~(\ref{eq_delta_response_function_result}) with the definitions of $r$ and $k_{0}$, Eqs.~(\ref{eq_definition_r}) and~(\ref{eq_definition_k0}), we finally obtain Eq.~(\ref{eq_doped_interband_response_function}) in the main text.

\section{Density-density response in the case of $p$-doping}
\label{appendix:appendix2}
Till now, all the results have been specified to the case of $n$ doping, when the Fermi energy $E_{\rm F}>0$ lies in conduction band. The response functions for the $p$-doped case ($E_{\rm F}<0$), however, can be obtained by slightly modifying the results which have been presented so far. 

In this Appendix, we show that the density response function of a $p$-doped LSM with Fermi wave vector $k_{\rm F}$ is equal to the response function of an $n$-doped LSM with the same Fermi wave number, provided, however, that one exchanges $m_{\rm c}$ with $m_{\rm v}$. 
This sounds natural, but it is not evident because of the different helicities of the valence- and conduction-band single-particle states. However, Eq.~(\ref{eq_form_factors_sum_intra}), Eq.~(\ref{eq_form_factors_sum_inter}) and the fact that the matrix $|D^{(\frac{3}{2})}_{\nu \nu'}\left(\theta\right)|^{2}$ is symmetric, imply that the form factors $\mathcal{A}_{\rm intra}(\theta)$ and $\mathcal{A}_{\rm inter}(\theta)$ which enter Eqs.~(\ref{eq_kubo_formula_intraband_contributions}) and~(\ref{eq_kubo_formula_interband_contributions}), would be the same if the helicity doublets ($\pm 1/2$ or $\pm 3/2$) corresponding to conduction and valence bands were exchanged. This property can be expressed more synthetically by introducing the notation:
\begin{equation}
{\cal A}_{\sigma, \sigma'}\left(\theta\right) \equiv \sum_{\nu \in \sigma} \sum_{\nu' \in \sigma'} \left|D^{(\frac{3}{2})}_{\nu \nu^{\prime}}(\theta)\right|^{2}~,
\end{equation}
where $\sigma$ and $\sigma'$ label bands ($+1$ corresponds to the conduction band and $-1$ to the valence band) and the sums run over helicity states $\nu$ and $\nu'$ belonging to the bands $\sigma$ and $\sigma'$. We have:
\begin{equation}
{\cal A}_{\sigma, \sigma'}(\theta) = \frac{1}{2}\left\{2 + \sigma \sigma'\left[3 \cos^{2}(\theta) -1\right]\right\} = {\cal A}_{-\sigma, -\sigma'}(\theta)~.
\end{equation}
An equivalent expression was obtained, with a different derivation, in Ref.~\onlinecite{tchoumakov_arxiv_2019}. The complete expression for the density-density response function of a non-interacting LSM, valid both in the $n$-doped and $p$-doped cases, reads as following:
\begin{equation} \label{eq:chi_general}
\chi^{(0)}_{nn}\left(\bm{q}, z\right) = \frac{1}{V} \sum_{\bm{k}, \sigma, \sigma'} \frac{n_{\bm{k}, \sigma}-n_{\bm{k}+\bm{q}, \sigma'}}{\hbar z + \epsilon_{\bm{k}, \sigma}-\epsilon_{\bm{k}+\bm{q}, \sigma'}} {\cal A}_{\sigma, \sigma'}(\theta_{\bm{k}, \bm{k}+\bm{q}})~,
\end{equation}
where $n_{{\bm k}, \sigma}$ denotes the Fermi occupation number at $T = 0$ associated with a single-particle state with wave vector ${\bm k}$ and band index $\sigma$, and $\epsilon_{{\bm k}, \sigma}$ is the single-particle energy
\begin{equation}
\epsilon_{{\bm k}, \sigma} = \begin{cases}
                             \hbar^{2} k^{2}/2m_{\rm c} & \text{for } \sigma = +1\\
                             - \hbar^{2} k^{2}/2m_{\rm v} & \text{for } \sigma = -1
                            \end{cases}~.
\end{equation}
Let us now consider a $p$-doped LSM with Fermi wave number $k_{\rm F}$, single-particle energies $\epsilon_{{\bm k}, \sigma}$, and occupation numbers
\begin{equation}
n_{{\bm k}, \sigma} = \begin{cases}
                             0 & \text{for } \sigma = +1\\
                             1 - \theta\left(k_{\rm F} - k\right) & \text{for } \sigma = -1
                            \end{cases}~,
\end{equation}
whose density response function is given by Eq.~(\ref{eq:chi_general}). Let us now look at an $n$-doped non-interacting LSM with the same Fermi wave number and {\it exchanged} valence and conduction effective masses. It would have single-particle energies $\tilde{\epsilon}_{{\bm k}, \sigma}$ and occupation numbers $\tilde{n}_{{\bm k}, \sigma}$, related to $\epsilon_{{\bm k}, \sigma}$ and $n_{{\bm k}, \sigma}$ by:
\begin{eqnarray}\label{eq:relation_pdoped_ndoped}
\tilde{\epsilon}_{\bm{k}, \sigma} &=& - \epsilon_{\bm{k}, -\sigma} \nonumber \\
\tilde{n}_{\bm{k}, \sigma} &=& 1 - n_{\bm{k}, -\sigma}~.
\end{eqnarray}
Its density-density response function would be:
\begin{equation}\label{eq:chi_pdoped}
\tilde{\chi}^{(0)}_{nn}\left(\bm{q}, z\right) = \frac{1}{V} \sum_{\bm{k}, \sigma, \sigma'} \frac{\tilde{n}_{\bm{k}, \sigma}-\tilde{n}_{\bm{k}+\bm{q}, \sigma'}}{\hbar z + \tilde{\epsilon}_{\bm{k}, \sigma}-\tilde{\epsilon}_{\bm{k}+\bm{q}, \sigma'}} {\cal A}_{\sigma, \sigma'}(\theta_{\bm{k}, \bm{k}+\bm{q}})~.
\end{equation}
Using Eq.~(\ref{eq:relation_pdoped_ndoped}) and relabeling dummy variables, we can rewrite Eq.~(\ref{eq:chi_pdoped}) as:
\begin{equation}
\begin{split}
 \tilde{\chi}^{(0)}_{nn}\left(\bm{q}, z\right)  & = \frac{1}{V}  \sum_{\bm{k}, \sigma, \sigma'} \Big[ \frac{n_{\bm{k}, \sigma}-n_{\bm{k}+\bm{q}, \sigma'}}{\hbar z + \epsilon_{\bm{k}, \sigma}-\epsilon_{\bm{k}+\bm{q}, \sigma'}} \\ &\times {\cal A}_{-\sigma', -\sigma}(\theta_{-\bm{k}-\bm{q}, -\bm{k}})\Big]~.
\end{split}
 \end{equation}
We finally observe that the form factors obey the following relations: ${\cal A}_{-\sigma',-\sigma}(\theta_{-\bm{k}-\bm{q}, -\bm{k}}) = {\cal A}_{\sigma',\sigma}(\theta_{-\bm{k}-\bm{q}, -\bm{k}}) = {\cal A}_{\sigma,\sigma'}(\theta_{\bm{k}, \bm{k}+\bm{q}})$. Replacing this result in Eq.~(\ref{eq_comparison_p_n_doped_results}) yields
\begin{equation}
\label{eq_comparison_p_n_doped_results}
\chi_{nn}^{(0)}(\bm{q},z) = \tilde{\chi}_{nn}^{(0)}(\bm{q},z)
\end{equation} 
and completes the proof.

\end{document}